\documentclass[12pt, a4paper]{article}
\usepackage[utf8]{inputenc}
\usepackage[left=2cm, right=2cm, bottom=2cm, top=2cm]{geometry}
\usepackage{xcolor}
\usepackage{svg}
\usepackage{booktabs}
\usepackage{graphicx}
\usepackage{caption}
\usepackage{subcaption}
\usepackage{amsmath, amsfonts, amssymb}
\usepackage{siunitx}
\DeclareMathOperator*{\argmax}{arg\,max}
\DeclareSIUnit{\angstrom}{\textup{\AA}}
\usepackage{mathtools}
\usepackage{csvsimple}
\usepackage{authblk}
\usepackage{enumitem}
\usepackage[autostyle=true]{csquotes}
\usepackage{xparse}
\usepackage{bm}
\usepackage{pgfkeys, pgfsys, pgfcalendar}  

\definecolor{amethyst}{rgb}{0.6, 0.4, 0.8}
\usepackage[colorlinks, citecolor=blue, linkcolor=teal, urlcolor=amethyst]{hyperref}
\usepackage[capitalise, nameinlink]{cleveref}

\newcommand*{\suppnameref}[1]{\hyperref[{#1}]{Supplementary material, \nameref*{#1}}}

\NewDocumentCommand{\qnameref}{sm}{%
  \enquote{%
    \IfBooleanTF{#1}{\nameref*{#2}}{\nameref{#2}}%
  }%
}
\NewDocumentCommand{\sfref}{sm}{\IfBooleanTF{#1}{\nameref*{#2}}{\nameref{#2}}%
}
\NewDocumentCommand{\suppqnameref}{sm}{%
  \enquote{%
    \IfBooleanTF{#1}{\suppnameref*{#2}}{\suppnameref{#2}}%
  }%
}

\newlist{steps}{enumerate}{10}
\setlist[steps]{label=\arabic*., ref=\arabic*}

\makeatletter
\if@cref@capitalise
\crefname{stepsi}{Step}{Steps}
\else
\crefname{stepsi}{step}{steps}
\fi
\makeatother

\usepackage[square, numbers, comma, sort&compress]{natbib}

\title{Pairing interacting protein sequences using masked language modeling}

\author{Umberto Lupo\textsuperscript{1,2,*,$\dagger$}, Damiano Sgarbossa\textsuperscript{1,2,*}, Anne-Florence Bitbol\textsuperscript{1,2,$\dagger$}}
\affil{\textbf{1} Institute of Bioengineering, School of Life Sciences, École Polytechnique Fédérale de Lausanne (EPFL), CH-1015 Lausanne, Switzerland\\
\textbf{2} SIB Swiss Institute of Bioinformatics, CH-1015 Lausanne, Switzerland\\
* These authors contributed equally to this work.\\
$^\dagger$ Emails: \href{mailto:umberto.lupo@epfl.ch}{\texttt{umberto.lupo@epfl.ch}}, \href{mailto:anne-florence.bitbol@epfl.ch}{\texttt{anne-florence.bitbol@epfl.ch}}}
\date{}

\begin{document}

\maketitle

\begin{abstract}
Predicting which proteins interact together from amino-acid sequences is an important task. We develop a method to pair interacting protein sequences which leverages the power of protein language models trained on multiple sequence alignments, such as MSA Transformer and the EvoFormer module of AlphaFold. We formulate the problem of pairing interacting partners among the paralogs of two protein families in a differentiable way. We introduce a method called DiffPALM that solves it by exploiting the ability of MSA Transformer to fill in masked amino acids in multiple sequence alignments using the surrounding context. MSA Transformer encodes coevolution between functionally or structurally coupled amino acids. We show that it captures inter-chain coevolution, while it was trained on single-chain data, which means that it can be used out-of-distribution. Relying on MSA Transformer without fine-tuning, DiffPALM outperforms existing coevolution-based pairing methods on difficult benchmarks of shallow multiple sequence alignments extracted from ubiquitous prokaryotic protein datasets. It also outperforms an alternative method based on a state-of-the-art protein language model trained on single sequences.  Paired alignments of interacting protein sequences are a crucial ingredient of supervised deep learning methods to predict the three-dimensional structure of protein complexes. DiffPALM substantially improves the structure prediction of some eukaryotic protein complexes by AlphaFold-Multimer, without significantly deteriorating any of those we tested. It also achieves competitive performance with using orthology-based pairing.
\end{abstract}

\section*{Significance statement}
Deep learning has brought major advances to the field of proteins. Self-supervised models, based on approaches from natural language processing and trained on large ensembles of protein sequences, efficiently learn statistical dependence in this data. This includes coevolution patterns between structurally or functionally coupled amino acids, which allows them to capture structural contacts. We propose a method to pair interacting protein sequences which leverages the power of a protein language model trained on multiple sequence alignments. Our method performs well for small datasets that are challenging for existing methods. It can improve structure prediction of protein complexes by supervised methods, which remains more challenging than that of single-chain proteins. 

\section*{Introduction}
\label{sec:intro}

Interacting proteins play key roles in cells, ensuring the specificity of signaling pathways and forming multi-protein complexes that act e.g.\ as molecular motors or receptors. Predicting protein-protein interactions and the structure of protein complexes are important questions in computational biology and biophysics.
Indeed, high-throughput experiments capable of resolving protein-protein interactions remain challenging~\cite{Rajagopala14}, even for model organisms, and experimental determination of protein complex structure is demanding.

A major advance in protein structure prediction was achieved by AlphaFold~\cite{Jumper21} and other deep learning approaches~\cite{Baek2021,Chowdhury21,Lin2022}. Extensions to protein complexes have been proposed~\cite{Humphreys2021,evans2021protein,mirdita2022colabfold,bryant2022improved}, including AlphaFold-Multimer (AFM)~\cite{evans2021protein}, but their performance is heterogeneous and less impressive than for monomers~\cite{Schweke23}. Importantly, the first step of AlphaFold is to build multiple-sequence alignments (MSAs) of homologs of the query protein sequence. The results of the CASP15 structure prediction contest demonstrated that MSA quality is crucial to further improving AlphaFold performance~\cite{Elofsson23,Alexander23}. For protein complexes involving several different chains (heteromers), paired MSAs, whose rows include actually interacting chains, can provide coevolution signal between interacting partners that is informative about inter-chain contacts~\cite{Weigt09,Bitbol16,Gueudre16,szurmant2018inter}. However, constructing paired MSAs poses the challenge of properly pairing sequences. Accordingly, the quality of pairings strongly impacts the accuracy of heteromer structure prediction \cite{bryant2022improved,Si22,Chen23}. Pairing interaction partners is difficult because many protein families contain several paralogous proteins encoded within the same genome. This problem is known as paralog matching. In prokaryotes, genomic proximity can often be used to solve it, since most interaction partners are encoded in close genomic locations \cite{orchard2013mintact, peters2016comprehensive}. However, this is not the case in eukaryotes. Large-scale coevolution studies of protein complexes~\cite{Ovchinnikov14,cong2019protein,Green21} and deep learning approaches~\cite{Zeng18,evans2021protein,Humphreys2021,bryant2022improved,mirdita2022colabfold} have paired sequences by using genomic proximity when possible~\cite{Ovchinnikov14,Zeng18,mirdita2022colabfold}, and/or by pairing together the closest, or equally ranked, hits to the query sequences, i.e.\ relying on approximate orthology~\cite{Zeng18,cong2019protein,Green21,evans2021protein,Humphreys2021,Pozzati21,bryant2022improved,mirdita2022colabfold}.

Aside from genomic proximity and orthology, phylogeny-based methods have addressed the paralog matching problem~\cite{goh2002co,Ramani03,Gertz03,Izarzugaza06,tillier2006codep,Izarzugaza08,Tillier09,Bradde10,Hajirasouliha12,ElKebir13}, exploiting similarities between evolutionary histories of interacting proteins \cite{pazos2001similarity,juan2008high,Jothi05,Ochoa10,Ochoa15}.
Other methods, based on coevolution~\cite{Casari95,Lapedes99,Weigt09,Morcos11,Marks11,Cheng14}, rely on correlations in amino-acid usage between interacting proteins~\cite{Burger08,Bitbol16,Gueudre16,Bitbol18}. These correlations arise from the need to maintain physico-chemical complementarity among amino acids in contact, and from shared evolutionary history~\cite{Marmier19,Gerardos22}. Phylogeny and coevolution can be explicitly combined, improving performance~\cite{Gandarilla23}.
However, coevolution-based approaches are data-thirsty and need large and diverse MSAs to perform well. This limits their applicability, especially to eukaryotic complex structure prediction. Nevertheless, the core idea of finding pairings that maximise coevolution signal holds promise for paralog matching and complex structure prediction.

We develop a new coevolution-based method for paralog matching which leverages recent neural protein language models taking MSAs as inputs~\cite{rao2021msa,Jumper21}. These models are one of the ingredients of the success of AlphaFold~\cite{Jumper21}. 
We focus on MSA Transformer~\cite{rao2021msa}, a protein language model which was trained on MSAs using the masked language modeling (MLM) objective in a self-supervised way. We introduce DiffPALM, a differentiable method for predicting paralog matchings using MLM. We show that it outperforms existing coevolution methods by a large margin on difficult benchmarks of shallow MSAs extracted from ubiquitous prokaryotic protein datasets. DiffPALM performance further quickly improves when known interacting pairs are provided as examples. Next, we apply DiffPALM to the hard problem of paralog matching for eukaryotic protein complexes. Among the complexes we tested, DiffPALM substantially improves structure prediction by AFM in some cases, and does not yield any significant deterioration. It also achieves competitive performance with using orthology-based pairing.

\section*{Results}

\subsection*{Leveraging MSA-based protein language models for paralog matching}\label{subsec:intro_diffpalm}

MSA-based protein language models, which include MSA Transformer~\cite{rao2021msa} and the EvoFormer module of AlphaFold~\cite{Jumper21}, are trained to correctly fill in masked amino acids in MSAs with the MLM loss (see \suppqnameref{supp_meth:MSA-Tr_MLM} and~\cite{rao2021msa,lupo2022protein} for details). To this end, they use the rest of the MSA as context, which allows them to capture coevolution. Indeed, MSA Transformer achieves state-of-the-art performance at unsupervised structural contact prediction~\cite{rao2021msa}, captures pairwise phylogenetic relationships between sequences~\cite{lupo2022protein}, and can be used to generate new sequences from given protein families~\cite{Sgarbossa23}.
While MSA Transformer was only trained on MSAs corresponding to single chains, inter-chain coevolution signal has strong similarities with intra-chain signal~\cite{Weigt09,Bitbol16,Gueudre16}. We find that MSA Transformer is able to detect inter-chain contacts from a properly paired MSA, while it cannot do so from a wrongly paired MSA, see \cref{fig-sup:concat_contacts}. Furthermore, we find that the MLM loss (used for the pre-training of MSA Transformer) decreases as the fraction of correctly matched sequences increases, see \cref{fig-sup:loss_vs_pairs}. These results demonstrate that MSA Transformer captures inter-chain coevolution signal. 

In this context, we ask the following question: Can we exploit MSA-based protein language models to address the paralog matching problem? Let us focus on the case where two MSAs have to be paired, which is the relevant one for heterodimers. Paralog matching amounts to pairing these two MSAs, each corresponding to one of two interacting protein families, so that correct interaction partners are placed on the same row of the paired MSA. Throughout, we will assume that interactions are one-to-one, excluding cross-talk, which is valid for proteins that interact specifically~\cite{Laub07}. Thus, within each species, assuming that there is the same number of sequences from both families, we aim to find the correct one-to-one matching that associates one protein from the first family to one protein from the second family. We also cover the case where the two protein families have different numbers of paralogs within the same species, see \qnameref{sec:methods}. Motivated by our finding that the MLM loss is lower for correctly paired MSAs than for incorrectly paired ones, we address the paralog matching problem by looking for pairings that minimise an MLM loss. A challenge is that the number of possible such one-to-one matchings scales factorially with the number of sequences in the species, making it difficult to find the permutation that minimises the loss by a brute-force search.  We address this challenge by formulating a differentiable optimization problem that can be solved using gradient methods, to yield configurations minimizing our MLM loss, see \qnameref{sec:methods}. We call our method \textbf{DiffPALM}, short for \textbf{Diff}erentiable \textbf{P}airing using \textbf{A}lignment-based \textbf{L}anguage \textbf{M}odels.

\subsection*{DiffPALM outperforms other coevolution methods on small MSAs}\label{subsec:results_prokaryotic}

We start out by considering a well-controlled benchmark dataset composed of ubiquitous prokaryotic proteins from two interacting families, namely histidine kinases (HKs) and response regulators (RRs)~\cite{Barakat09,Barakat11}, see \suppqnameref{subsec:Datasets}. These proteins interact together within prokaryotic two-component signaling systems, important pathways that enable bacteria to sense and respond to environment signals~\cite{Laub07}. They possess multiple paralogs (on the order of ten per genome, with substantial variability), and have strong specificity for their cognate partners. Because most cognate HK-RR pairs are encoded in the same operon, many interaction partners are known from genome proximity, which enables us to assess performance. In addition, earlier coevolution methods for paralog matching were tested on this dataset, allowing rigorous comparison~\cite{Bitbol16,Bitbol18,Gandarilla23}. Here, we focus on datasets comprising about 50 cognate HK-RR pairs. Indeed, this small data regime is problematic for existing coevolution methods, which require considerably deeper alignments to achieve good performance \cite{Bitbol16,Gueudre16,Bitbol18,Gandarilla23}. Furthermore, this regime is highly relevant for eukaryotic complexes, because their homologs have relatively small sequence diversity, as shown by the effective depth of their MSAs in \cref{tab:dataset_pdb}. While prokaryotic proteins such as HKs and RRs feature large diversity, focusing on small datasets allows us to address the relevant regime of low diversity in this well-controlled benchmark case. We hypothesize that MSA Transformer's extensive pre-training can help to capture coevolution even in these difficult cases.
To assess this, we first test two variants of our DiffPALM method (see \qnameref{sec:methods}) on $40$ MSAs from the HK-RR dataset comprising about 50 HK-RR pairs each (see \suppqnameref{subsec:Datasets}). We first address the \textit{de novo} pairing prediction task, starting from no known HK-RR pair, and then we study the impact of starting from known pairs.

\begin{figure}[h!]
\centering
\includegraphics[width=0.7\textwidth]{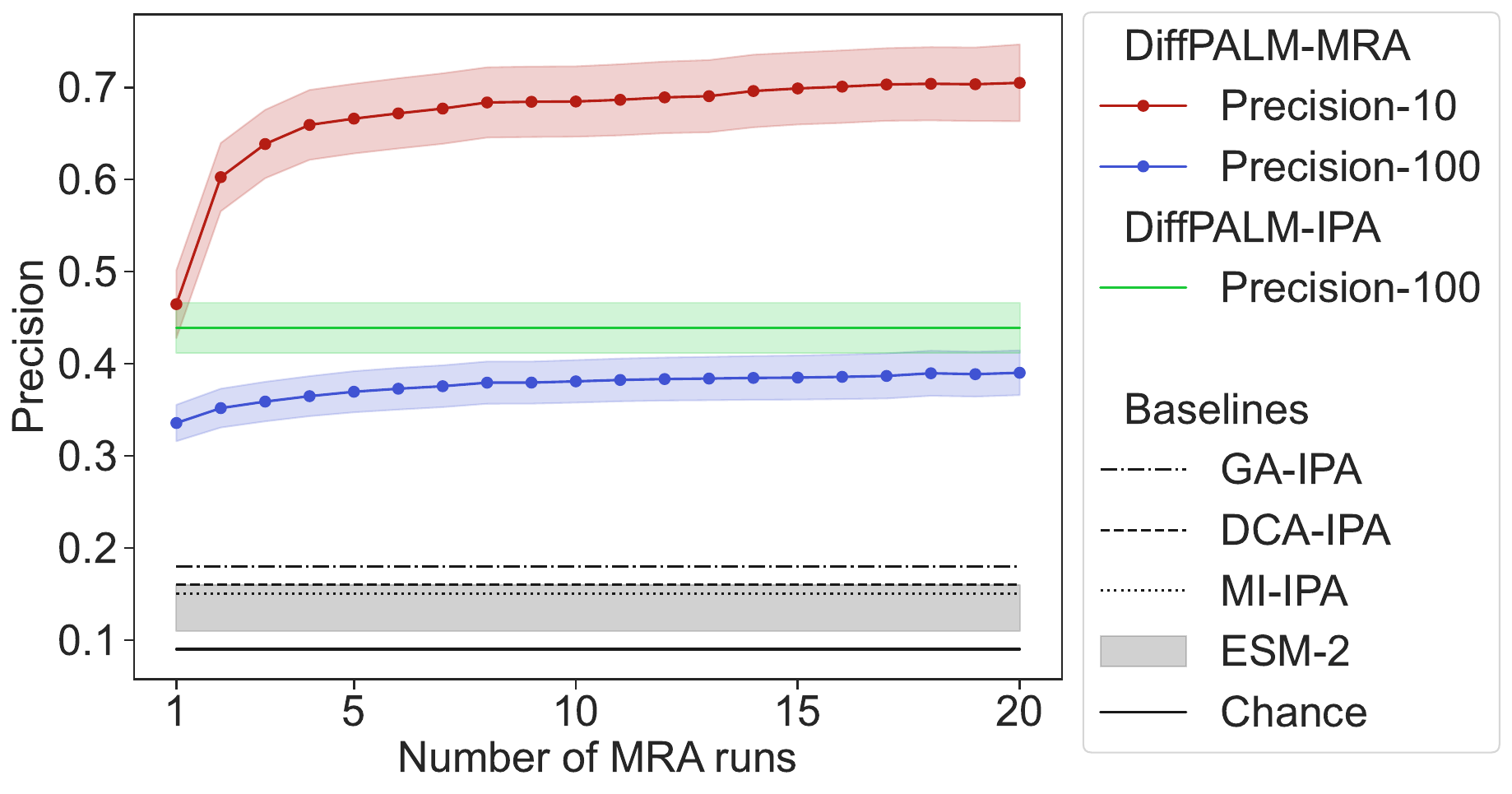}
\caption{\textbf{Performance of DiffPALM on small HK-RR MSAs.} The performance of two variants of DiffPALM (MRA and IPA, see \qnameref{subsec:MRA-IPA}) is shown versus the number of runs used for the MRA variant, for $40$ MSAs comprising about 50 HK-RR pairs. The chance expectation, and the performance of various other methods, are reported as baselines. Three existing coevolution-based methods are considered: DCA-IPA \cite{Bitbol16}, MI-IPA \cite{Bitbol18}, and GA-IPA \cite{Gandarilla23}. We also consider a pairing method based on the scores given by the ESM-2 (650M) single-sequence protein language model \cite{Lin2022}, see \qnameref{subsec:methods_esm2}. With all methods, a full one-to-one within-species pairing is produced, and performance is measured by precision (also called positive predictive value or PPV), namely, the fraction of correct pairs among predicted pairs. The default score is ``precision-100'', where this fraction is computed over all predicted pairs (100\% of them). For DiffPALM-MRA, we also report ``precision-10'', which is calculated over the top $10 \%$ predicted pairs, when ranked by predicted confidence within each MSA (see \qnameref{sec:methods}). For DiffPALM, we plot the mean performance on all MSAs (color shading), and the standard error range (shaded region). For our ESM-2 based method, we consider 10 different values of masking probability $p$ from $0.1$ to $1.0$, and we report the range of precisions obtained (gray shading). For other baselines, we report the mean performance on all MSAs.}
\label{fig:results_mra}
\end{figure}

\cref{fig:results_mra} shows that DiffPALM performs better than the chance expectation, obtained for random within-species matching. Moreover, it outperforms other coevolution-based methods, namely DCA-IPA \cite{Bitbol16}, MI-IPA \cite{Bitbol18}, which rely respectively on Potts models and on mutual information, and GA-IPA \cite{Gandarilla23}, which combines these coevolution measures with sequence similarity, a proxy for phylogeny. Importantly, these results are obtained without giving any paired sequences as input to the algorithm. The performance of DiffPALM is particularly good for pairs with high confidence score (see \qnameref{subsec:confidence}), as shown by the ``precision-10'' curve, which focuses on top $10 \%$ predicted pairs, when ranked by predicted confidence (see \cref{fig:results_mra}). We also propose a method based on a protein language model trained on single sequences, ESM-2 (650M)~\cite{Lin2022}, see \qnameref{subsec:methods_esm2}. DiffPALM also outperforms this method, even though the latter is faster (no need for backpropagation) and is formulated as a linear matching problem, which is solved exactly. This confirms that the coevolution information contained in the MSA plays a key role in the performance of DiffPALM, which is based on MSA Transformer. A key strength of MSA Transformer and thus of DiffPALM is that they leverage the power of large language models while starting from MSAs, and thus allow direct access to the coevolution signal. \cref{fig:results_mra} shows that both variants of DiffPALM, namely MRA and IPA (see \qnameref{sec:methods}) outperform all baselines, and that precision of MRA increases with the number of runs used (see \cref{tab:all-data} for details).
Note that the distribution of precision-10 scores over the different MSAs we considered is skewed, especially after many MRA runs, see figure \cref{fig-sup:results_distributions}. For many MSAs, almost perfect scores are reached, while performance is bad for a few others. MSAs with smaller average number of sequence per species tend to yield larger precision, as the pairing task is then easier.

\newpage

\begin{figure}[h!]
\centering
\includegraphics[width=0.7\textwidth]{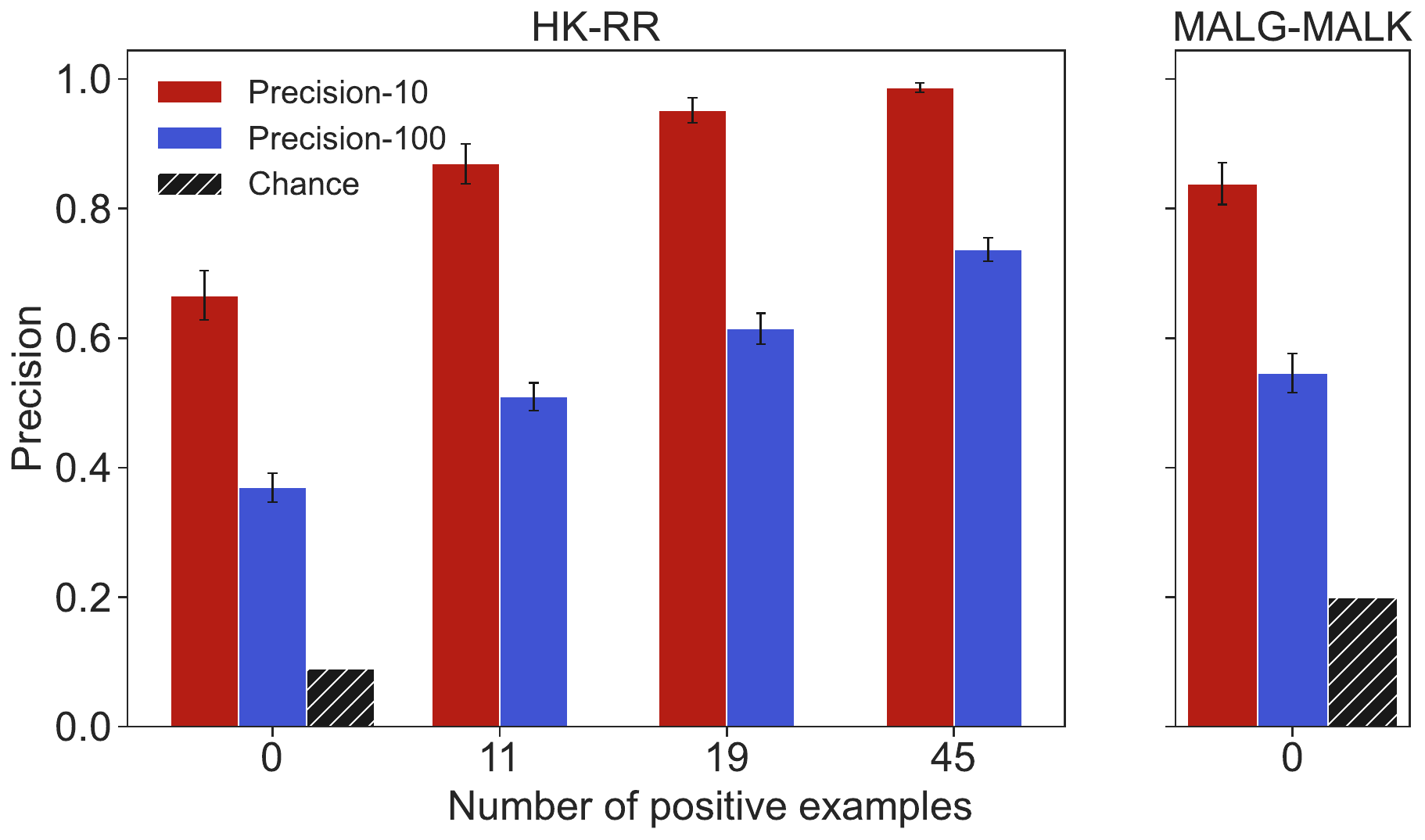}
\caption{\textbf{Impact of positive examples and extension to another pair of protein families.} We report the performance of DiffPALM with 5 MRA runs (measured as precision-100 and precision-10, see \cref{fig:results_mra}), for various numbers of positive examples, on the same HK-RR MSAs as in \cref{fig:results_mra} (left panel). We also report the performance of DiffPALM for similarly-sized MALG-MALK MSAs (right panel). In both cases, we show the mean value over the 40 different MSAs with its standard error interval, and we plot the chance expectation for reference.} 
\label{fig:results_bars}
\end{figure}

So far, we addressed \textit{de novo} pairing prediction, where no known HK-RR pair is given as input. Can DiffPALM precision increase by exploiting ``positive examples'' of known interacting partners? This is an important question, since experiments on model species may for instance give some positive examples (see \qnameref{subsec:paralog_matching_problem}). To address it, we included different numbers of positive examples, by using the corresponding non-masked interacting pairs as context (see \qnameref{subsec:MLM_loss_meth}). The left panel of \cref{fig:results_bars} shows that the performance of DiffPALM significantly increases with the number of positive examples used, reaching almost perfect performance for the highest-confidence pairs (precision-10).

While we focused on HK-RR pairing so far, DiffPALM is a general method. To assess how it extends to other cases, we consider another pair of ubiquitous prokaryotic proteins, namely homologs of the \textit{E. coli} proteins MALG-MALK, which are involved in ABC transporter complexes. These proteins form permanent complexes, while HK-RR interact transiently to transmit signal. The right panel of \cref{fig:results_bars} shows results obtained on $40$ MSAs comprising about 50 MALG-MALK pairs, without positive examples. We observe that DiffPALM outperforms the chance expectation by a large margin. It also significantly outperforms existing coevolution methods~\cite{Bitbol16,Bitbol18,Gandarilla23}, as well as our method based on ESM-2 (650M), see \cref{tab:all-data}. Note that all approaches yield better performance for MALG-MALK than for HK-RR, as the number of MALG-MALK pairs per species is smaller than that of HK-RR pairs.
Finally, while \cref{fig:results_bars} reports the final MRA performance, \cref{fig-sup:results_iters} shows that both performance scores increase with the number of MRA runs in all cases.

\subsection*{Using DiffPALM for eukaryotic complex structure prediction by AFM}\label{subsec:results_eukaryotic}

An important and more challenging application of DiffPALM is predicting interacting partners among the paralogs of two families in eukaryotic species. Indeed, eukaryotes often have many paralogs per species \cite{Makarova05} but eukaryotic-specific protein families generally have fewer total homologs and smaller diversity than prokaryotes. Moreover, most interacting proteins are not encoded in close proximity in eukaryotic genomes.
Paired MSAs are a key ingredient of protein complex structure prediction by AFM~\cite{evans2021protein,bryant2022improved}. When presented with query sequences, the default AFM pipeline \cite{evans2021protein} retrieves homologs of each of the chains. Within each species, homologs of different chains are ranked according to Hamming distance to the corresponding query sequence. Then, equal-rank sequences are paired. Can DiffPALM improve complex structure prediction by AFM? To address this question, we consider 15 complexes, listed in \cref{tab:dataset_pdb}, whose structures are not included in the training set of the AFM release we used (see \suppqnameref{subsec:generalities_AFM}), and for which the default AFM complex prediction was reported to perform poorly~\cite{Chen23,evans2021protein}  (see \suppqnameref{supp_datasets_eukaryotic}).

\begin{figure}[h!]
\centering
\includegraphics[width=\textwidth]{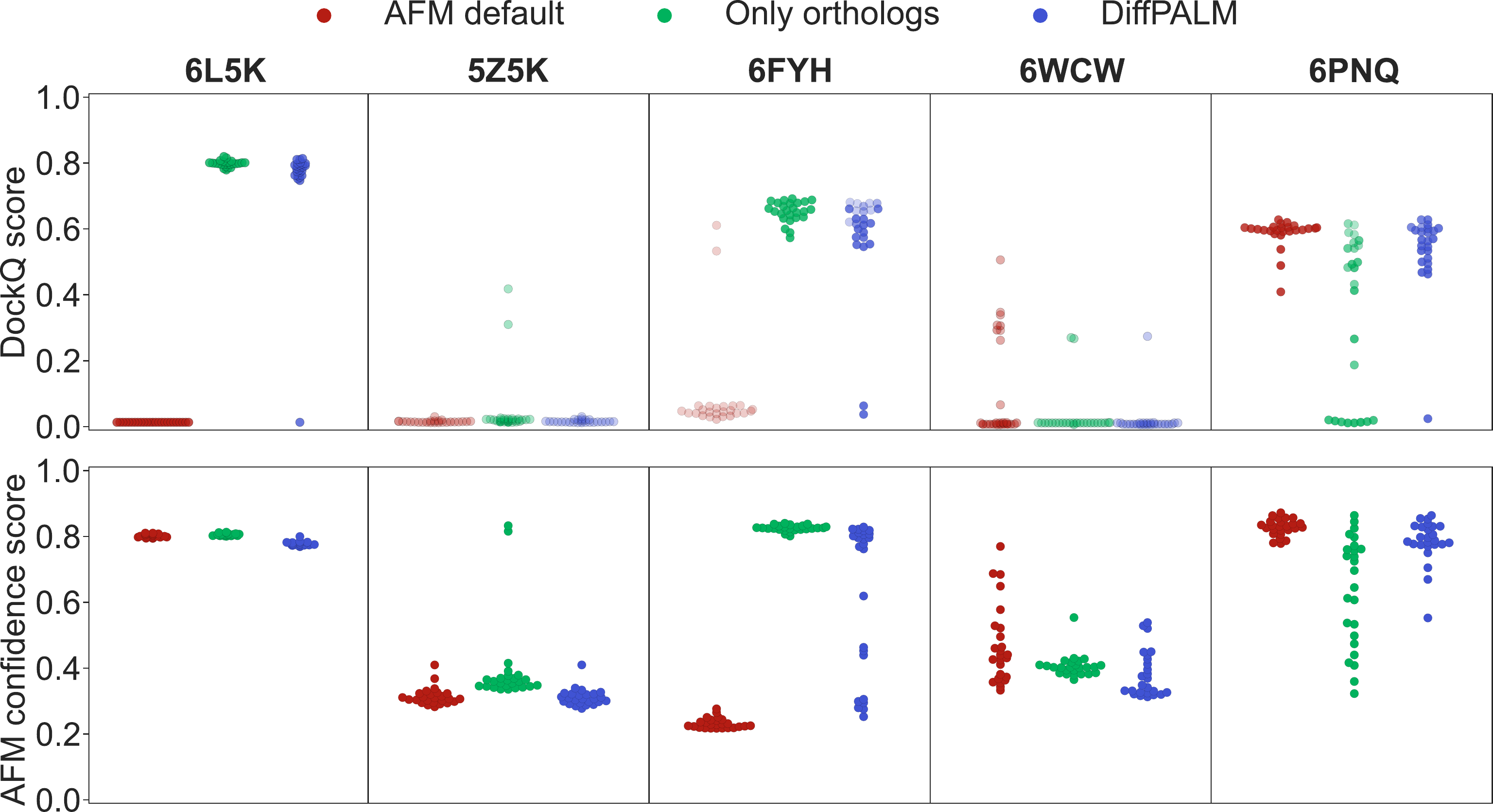}
\caption{\textbf{Performance of AFM using different pairing methods.} We use AFM to predict the structure of protein complexes starting from differently paired MSAs, each of them constructed from the same initial unpaired MSAs. Three pairing methods are considered: the default one of AFM, only pairing orthologs to the two query sequences, and a single run of DiffPALM (equivalent to one MRA run). Performance is evaluated using DockQ scores (top panels), a widely used measure of quality for protein-protein docking~\cite{Basu16}, and the AFM confidence scores (bottom panels), see \suppqnameref{subsec:generalities_AFM}. The latter are also used as transparency levels in the top panels, where more transparent markers denote predicted structures with low AFM confidence. For each query complex, AFM is run five times.  Each run yields 25 predictions which are ranked by AFM confidence score. The five top predicted structures are selected from each run, giving 25 predicted structures in total for each complex.
Out of the 15 complexes listed in \cref{tab:dataset_pdb}, we restrict to those where any two of these three pairing methods yield a significant difference ($>0.1$) in average DockQ scores for at least one set of predictions coming from different runs but with the same within-run rank according to AFM confidence. Panels are ordered by increasing mean DockQ score for the AFM default method.
} 
\label{fig:results_afm_swarm}
\end{figure}

\cref{fig:results_afm_swarm} shows that DiffPALM can improve complex structure prediction by AFM (see \cref{fig-sup:results_trajectories} for details). This suggests that it is able to produce better paired MSAs than those from the default AFM pipeline. In particular, substantial improvements are obtained for the complexes with PDB identifiers 6L5K and 6FYH, see \cref{fig-sup:6l5k_pymol,fig-sup:6fyh_pymol} for structural visualizations. Among the complexes we considered, 6L5K and 6FYH have large effective (pairable) MSA depths, see \cref{tab:dataset_pdb}. Conversely, complexes with very small raw or effective MSA depths do not significantly benefit from DiffPALM. Thus, DiffPALM is sensitive to MSA depth and diversity, albeit having less stringent such requirements than other coevolution methods. In most cases, the quality of structures predicted using DiffPALM pairing is comparable to that obtained using the pairing method adopted e.g.\ by ColabFold~\cite{mirdita2022colabfold}, where only the orthologs of the two query sequences, identified as their best hits, are paired in each species (resulting in at most one pair per species)~\cite{Zeng18,cong2019protein,Green21,Pozzati21,bryant2022improved,mirdita2022colabfold}, see \cref{fig:results_afm_swarm}. Note however that, for 6PNQ, the ortholog-only pairing method is outperformed both by DiffPALM and by AFM default. Indeed, the raw and effective MSA depths are smaller for this structure than e.g.\ for 6L5K and 6FYH (see \cref{tab:dataset_pdb}). Thus, further reducing diversity by restricting to paired orthologs may be negatively impacting structure prediction in this case. Given the good results obtained overall with orthology-based pairings, we tried using them as positive examples for DiffPALM. Given the very good precision obtained by DiffPALM for high-confidence HK-RR pairs (see above), we also tried restricting to high-confidence pairs. For most structures, we obtained no significant improvement over the standard DiffPALM using these variants (see \cref{fig-sup:results_afm_swarm}). However, for 6WCW, we could generate several higher-quality structures, particularly when using orthologs as positive examples.

Although DiffPALM achieves similar performance on these structure prediction tasks as using orthology, it predicts some pairs that are quite different from orthology-based pairs. 
Indeed, \cref{fig-sup:results_conf} shows that the fraction of pairs identically matched by DiffPALM and by orthology is often smaller than $0.5$.
\cref{fig-sup:results_ham} further shows that, for the sequences that are paired differently by DiffPALM and by orthology, the Hamming distances between the two predicted partners is often above $0.5$. Nevertheless, most of the pairs that are predicted both by DiffPALM and by using orthology have high DiffPALM confidence (see \cref{fig-sup:results_conf}), confirming the importance of these pairs.

\section*{Discussion}

We developed DiffPALM, a method for pairing interacting protein sequences that builds on MSA Transformer~\cite{rao2021msa}, a protein language model trained on MSAs. MSA Transformer efficiently captures coevolution between amino acids, thanks to its training to fill in masked amino acids using the surrounding MSA context~\cite{rao2021msa,lupo2022protein,Sgarbossa23}. We showed that it also captures inter-chain coevolutionary signal, despite being trained on single-chain MSAs. We leveraged this ability in DiffPALM by using a masked language modeling loss as a coevolution score and looking for the pairing that minimizes it. We formulated the pairing problem in a differentiable way, allowing us to use gradient methods. On shallow MSAs extracted from controlled prokaryotic benchmark datasets, DiffPALM outperforms existing coevolution-based methods as well as a method based on a state-of-the-art language model trained on single sequences. Its performance quickly increases when adding examples of known interacting sequences. Paired MSAs of interacting partners are a key ingredient to complex structure prediction by AFM. We found that using DiffPALM can improve the performance of AFM, and achieves competitive performance with orthology-based pairing.

Recent work \cite{Chen23} also used MSA Transformer for paralog matching, in a method called ESMPair. It relies on column attention matrices and compares them across the MSAs of interacting partners. This makes it quite different from DiffPALM, which relies on coevolutionary information via the MLM loss. ESMPair may be more closely related to phylogeny-based~\cite{Si22} or orthology-based pairing methods, since column attention encodes phylogenetic relationships \cite{lupo2022protein}. 13 out of the 15 eukaryotic protein complexes we considered were also studied in \cite{Chen23}, but no substantial improvement (and often a degradation of performance) was reported for those when using ESMPair instead of the default AFM pairing, except for 7BQU. By contrast, DiffPALM yields strong improvements for 6L5K and 6FYH, and no significant performance degradation. Explicitly combining coevolution and phylogeny using MSA Transformer is a promising direction to further improve partner pairing. Indeed, such an approach has already improved traditional coevolution methods \cite{Gandarilla23}. Other ways of improving MSA usage by AFM have also been proposed~\cite{Bryant23} and could be combined with advances in pairing. Besides improving MSA construction~\cite{ZhengAbstract} and the extraction of MSA information, other promising approaches include exploiting structural alignments~\cite{Liu23}, using massive sampling and dropout~\cite{Wallner23}, and combining AFM with more traditional docking methods~\cite{Ghani21,Olechnovic23}, which has allowed e.g.\ to improve structure prediction of 6A6I~\cite{Ghani21}. 

DiffPALM illustrates the power of neural protein language models trained on MSAs, and their ability to capture the rich structure of biological sequence data. The fact that these models encode inter-chain coevolution, while they are trained on single-chain data, shows their ability to generalize. We used MSA Transformer in a zero-shot setting, without fine-tuning it to the task of interaction partner prediction. Such fine-tuning could yield further performance gains \cite{Hawkins22}. 

The fact that DiffPALM outperforms existing coevolution methods~\cite{Bitbol16,Bitbol18,Gandarilla23} for shallow MSAs is reminiscent of the impressive performance of MSA Transformer at predicting structural contacts from shallow MSAs~\cite{rao2021msa}. While traditional coevolution methods either compute local coevolution scores for two columns of an MSA~\cite{Bitbol18} or build a global model for an MSA~\cite{Bitbol16,Gueudre16}, MSA Transformer was trained on large ensembles of MSAs and shares parameters across them. This presumably allows it to transfer knowledge between MSAs, and to bypass the usual needs for deep MSAs of traditional coevolution methods~\cite{Burger08,Bitbol16,Gueudre16,Bitbol18,Gandarilla23}, or of MSA-specific transformer models~\cite{Meynard-Piganeau23}. This constitutes major progress for the use of coevolution signal. 

After the transformative progress brought by deep learning to protein structure prediction~\cite{Jumper21,Baek2021,Chowdhury21,Lin2022}, predicting protein complex structure and ligand binding sites is fast advancing with AFM and related methods, but also with other deep learning models based on structural representations~\cite{Gainza2020,Tubiana22,Krapp23,Pun22}. Combining the latter with the power of sequence-based language models may bring even further progress.

\section*{Methods}
\label{sec:methods}

\subsection*{The paralog matching problem}\label{subsec:paralog_matching_problem}

\paragraph{Goal and notations.} Paralog matching amounts to pairing a pair of MSAs, each one corresponding to one of the two protein families considered. We assume that interactions are one-to-one. Let $\mathcal{M}^{(\mathrm{A})}$ and $\mathcal{M}^{(\mathrm{B})}$ be the (single-chain) MSAs of two interacting protein families A and B, and let $K$ denote the number of species represented in both MSAs and comprising more than one unmatched sequence in at least one MSA. Species represented in only one MSA are discarded since no within-species matching is possible for them. Species with only one unmatched sequence in each MSA are not considered further since pairing is trivial. There may also be $N_{\mathrm{pos}}$ known interacting pairs: they are treated separately, as positive examples (see below). Here we focus on the unmatched sequences.
For $k = 1, \ldots, K$, let $N^{(\mathrm{A})}_k$ and $N^{(\mathrm{B})}_k$ denote the number of unmatched sequences belonging to species $k$ in $\mathcal{M}^{(\mathrm{A})}$ and $\mathcal{M}^{(\mathrm{B})}$ (respectively).

\paragraph{Dealing with asymmetric cases.} The two protein families considered may have different numbers of paralogs within the same species. Assume, without loss of generality, that $N^{(\mathrm{A})}_k < N^{(\mathrm{B})}_k$ for a given $k$. To solve the matching problem with one-to-one interactions, we would like to pick, for each of the $N^{(\mathrm{A})}_k$ sequences in $\mathcal{M}^{(\mathrm{A})}$, a single and exclusive interaction partner out of the $N^{(\mathrm{B})}_k$ available sequences in $\mathcal{M}^{(\mathrm{B})}$. The remaining sequences of the species in $\mathcal{M}^{(\mathrm{B})}$ are left unpaired.
In practice, we achieve this by augmenting the original set of species-$k$ sequences from $\mathcal{M}^{(\mathrm{A})}$ with $N^{(\mathrm{B})}_k - N^{(\mathrm{A})}_k$ ``padding sequences'' made entirely of gap symbols.
By doing so (and analogously when $N^{(\mathrm{A})}_k > N^{(\mathrm{B})}_k$), the thus-augmented interacting MSAs have the same number $N_k := \max(N^{(\mathrm{A})}_k, N^{(\mathrm{B})}_k)$ of sequences from each species $k$.
In practice, this method is used for the AFM complex structure prediction, while the curated benchmark prokaryotic MSAs do not have asymmetries (see \suppqnameref{subsec:Datasets}).

\paragraph{Formalization.} The paralog matching problem corresponds to finding, within each species $k$, a mapping that associates one sequence of $\mathcal{M}^{(\mathrm{A})}$ to one sequence of $\mathcal{M}^{(\mathrm{B})}$ (and reciprocally). Thus, within each species $k$, one-to-one matchings can be encoded as permutation matrices of size $N_k \times N_k$. A brute-force search through all possible within-species one-to-one matchings would scale factorially with the size $N_k$ of each species, making it prohibitive. Note that the Iterative Pairing Algorithm (IPA)~\cite{Bitbol16,Bitbol18} is an approximate method to solve this problem when optimizing coevolution scores. Here, we introduce another one, which allows to leverage the power of deep learning.

\paragraph*{Exploiting known interacting partners.}
Our use of a language model allows for contextual conditioning, a common technique in natural language processing.
Indeed, if any correctly paired sequences are already known, they can be included as part of the joint MSA input to MSA Transformer.
In this case, we exclude their pairing from the optimization process -- in particular, by not masking any of their amino acids, see below.
We call these known paired sequences ``positive examples''. In \cref{fig:results_bars}, we randomly sampled species and included all their pairs as positive examples, until we reached the desired depth $N_{\mathrm{pos}}\pm 10\%$.
For eukaryotic complex structure prediction, we treated the query sequence pair as a positive example.

\subsection*{DiffPALM: Paralog matching based on MLM}
\label{subsec:algo}

\begin{figure}[!h]
    \centering
    \includegraphics[width=\textwidth]{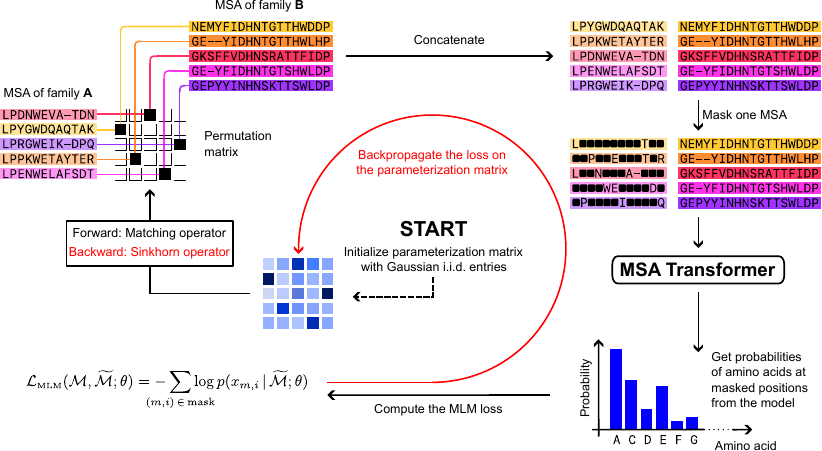}
    \caption{\textbf{Schematic of the DiffPALM method.} First, the parameterization matrices $X_k$ are initialized, and then the following steps are repeated until the loss converges: (1) Compute the permutation matrix $M(X_k)$ and use it to shuffle $\mathcal{M}^{(\mathrm{A})}$ relative to $\mathcal{M}^{(\mathrm{B})}$. Then pair the two MSAs. (2) Randomly mask some tokens of one of the two sides of the paired MSA and compute the MLM loss \cref{eq:MLM_loss}. (3) Backpropagate the loss and update the parameterization matrices $X_k$, using the Sinkhorn operator $\hat{S}$ for the backward step instead of the matching operator $M$ (see \suppqnameref{supp_meth:diff_matching}).}
    \label{fig:pipeline}
\end{figure}

Here, we explain our paralog matching method based on MLM, which we call DiffPALM.  Background information on MSA Transformer and its MLM loss is collected in \suppqnameref{supp_meth:MSA-Tr_MLM}. DiffPALM exploits our differentiable framework for optimizing matchings, see \suppqnameref{supp_meth:diff_matching}. The key steps are summarized in \cref{fig:pipeline}.

\paragraph{Construction of an appropriate MLM loss. \label{subsec:MLM_loss_meth}} Using the tools just described, we consider two interacting MSAs (possibly augmented with padding sequences), still denoted by $\mathcal{M}^{(\mathrm{A})}$ and $\mathcal{M}^{(\mathrm{B})}$. Given species indexed by $k = 1, \ldots, K$, we initialize a set $\{X_k\}_{k = 1, \ldots, K}$ of square matrices of size $N_k \times N_k$ (the case $K = 1$ corresponds to $X$ in \suppqnameref{supp_meth:diff_matching}). We call these ``parameterization matrices''. By applying to them the matching operator $M$ [see \cref{eq:matching_op}], we obtain the permutation matrices $\{M(X_k)\}_{k = 1, \ldots, K}$, encoding matchings within each species in the paired MSA. Using gradient methods, we optimize the parameterization matrices so that the corresponding permutation matrices yield a paired MSA with low MLM loss.
More precisely, paired MSAs are represented as concatenated MSAs with interacting partners placed on the same row,\footnote{We employ no special tokens to demarcate the boundary between one sequence and its partner.} and our MLM loss for this optimization is computed as follows: 
\begin{enumerate}[noitemsep,topsep=0pt]
    \item Perform a shuffle of $\mathcal{M}^{(\mathrm{A})}$ relative to $\mathcal{M}^{(\mathrm{B})}$ using the permutation matrix $M(X_k)$ in each species $k$ (plus an optional noise term, see below), to obtain a paired MSA $\mathcal{M}$; 
    \item Generate a mask for $\mathcal{M}$ (excluding any positive example tokens from the masking); 
    \item Compute MSA Transformer's MLM loss for that mask, see \cref{eq:MLM_loss}.
\end{enumerate}
Importantly, we only mask tokens from one of the two MSAs, chosen uniformly at random within that MSA with a high masking probability $p = 0.7$.\footnote{In contrast, uniformly random masking with $p = 0.15$ was used during MSA Transformer's pre-training \cite{rao2021msa}.}
Our rationale for using large masking probabilities is that it forces the model to predict masked residues in one of the two MSAs by using information coming mostly from the other MSA -- see \cref{fig-sup:loss_vs_pairs}.
We stress that, if padding sequences consisting entirely of gaps are present (see above), we mask these symbols with the same probability as those coming from ordinary sequences.
Of the two MSAs to pair, we mask the one with shorter length if no padding sequences exist (i.e.\ here for our prokaryotic benchmark datasets).
Else, if lengths are comparable but one MSA contains considerably more padding sequences than the other, we preferentially mask that MSA.
Otherwise, we randomly choose which of the two MSAs to mask.

We fine-tuned all the hyperparameters involved in our algorithm using two joint MSAs of depth $\sim 50$, constructed by selecting random species from the HK-RR dataset (see \suppqnameref{subsec:Datasets}).

\paragraph{Noise and regularization.} Following \cite{Mena2018}, after updating (or initializing) each $X_k$, we add to it a noise term given by a matrix of standard i.i.d.\ Gumbel noise multiplied by a scale factor.
The addition of noise ensures that the $X_k$ do not get stuck at degenerate values for the right-hand side of \cref{eq:matching_op}, and more generally may encourage the algorithm to explore larger regions in the space of permutations.
As scale factor for this noise we choose $0.1$ times the sample standard deviation of the current entries of $X_k$, times a global factor tied to the optimizer scheduler (see next paragraph).
Finally, since the matching operator is scale-invariant, we can regularize the matrices $X_k$ to have small Frobenius norm.
We find this to be beneficial and implement it through weight decay, set to be $w = 0.1$.

\paragraph{Optimization.} We backpropagate the MLM loss on the parameterization matrices $X_k$. We use the AdaDelta optimizer \cite{Zeiler12} with an initial learning rate $\gamma = 9$ and a ``reduce on loss plateau'' learning rate scheduler which decreases the learning rate by a factor of $0.8$ if the loss has not decreased for more than $20$ gradient steps after the learning rate was last set. The learning rate scheduler also provides the global scale factor which, together with the standard deviation of the entries of $X_k$, dynamically determines the magnitude of the Gumbel noise (see previous paragraph).

\paragraph{Exploring the loss landscape through multiple initializations.} \label{subsec:mult_init}
We observe that the initial choice of the parameterization set $\{X_k\}_{k = 1, \ldots, K}$ strongly impact results. Slightly different initial conditions for $X_k$ lead to very different final permutation matrices. Furthermore, we observe fast decrease in the loss when the $X_k$ are initialized to be exactly zero (our use of Gumbel noise means that we break ties randomly when computing the permutation matrices $M(X_k)$; if noise is not used, similar results can be achieved by initializing $X_k$ with entries very close to zero). Thus, we can cheaply probe different paths in the loss landscape by performing several short runs using zero-initialized parameterization matrices $X_k$. In practice, we use 20 different such short runs each consisting of 20 gradient steps. Then, we average all the final parameterizations together to warm-start a longer run made up of 400 gradient steps.

\paragraph{Result and confidence.} \label{subsec:confidence} We observe that, even though the loss generally converges to a minimum average value during our optimization runs, there are often several distinct hard permutations associated to the smallest loss values.
This may indicate a flattening of the loss landscape relative to the inherent fluctuations in the MLM loss, and/or the existence of multiple local minima.
To extract a single matching per species from one of our runs (or indeed from several runs, see \qnameref{subsec:MRA-IPA}), we average the hard permutation matrices associated to the $q$ lowest losses, and evaluate the matching operator [\cref{eq:matching_op}] on the resulting averages. We find final precision-100 figures to be quite robust to the choice of $q$. On the other hand, for individual (warm-started) runs as described in \qnameref{subsec:mult_init}, precision-10 benefits from setting $q$ to its maximum possible value of $400$.

Furthermore, we propose using each entry in the averaged permutation matrices as an indicator of the model's confidence in the matching of the corresponding pair of sequences.
Indeed, pairs that are present in most low-loss configurations are presumably essential for the optimization process and are captured most of the times, pushing their confidence value close to 1.
Conversely, non-interacting pairs are in most of the cases associated to higher losses and therefore appear sporadically, obtaining confidences close to zero.
Accordingly, we refer to the averaged hard permutations used to extract a single matching per species as ``confidence matrices'', and to the final in-species matchings as ``consensus permutations''.

\paragraph{Improving precision: MRA and IPA.} \label{subsec:MRA-IPA}
We propose two methods for improving precision further.
In the first method, which we call Multi-Run Aggregation (MRA), we perform $N_\mathrm{runs}$ independent optimization runs for each interacting MSA.
Then, we collect together the hard permutations independently obtained from each run, and aggregate the $q = 400$ lowest-loss permutations from this larger collection to obtain more reliable confidence matrices and hard permutations.

The second method is an iterative procedure analogous to the Iterative Pairing Algorithm (IPA) of Refs.~\cite{Bitbol16,Bitbol18}, and named after it. The idea is to gradually add pairs with high confidence as positive examples.
Assuming a paired MSA containing a single species for notational simplicity, the $n$-th iteration (starting at $n = 1$) involves the following steps:
\begin{enumerate}[noitemsep,topsep=0pt]
    \item Perform an optimization run and extract from it a confidence matrix $C^{(n)}$ as described in \qnameref{subsec:confidence}, using the currently available positive examples;
    \item Compute the moving average $\tilde{C}^{(n)} = \mathrm{mean}(C^{(n)}, \tilde{C}^{(n - 1)}, \ldots, \tilde{C}^{(1)})$ (where $\tilde{C}^{(1)} \equiv C^{(1)}$); 
    \item Define candidate matchings via the consensus permutation $M^{(n)} = M(\tilde{C}^{(n)})$;
    \item Repeat Steps 1-3 a maximum of 3 times, until the average MLM loss estimated using $M^{(n)}$, and $200$ random masks, is lower or statistically insignificantly higher\footnote{\label{ftnt_ttest}Based on a two-sample T-test: we say ``statistically insignificantly'' when $p \geq 0.95$, and ``statistically significantly'' when $p < 0.05$.} than what could have been obtained using $M^{(n - 1)}$ and the same positive examples as in Step 1; 
    \item If Step 4 fails, set $\tilde{C}^{(n)} = \tilde{C}^{(n - 1)}$ and $M^{(n)} = M(\tilde{C}^{(n - 1)})$ (but removing rows and columns corresponding to the positive examples added at iteration $n - 1$);
    \item Check that the average MLM loss estimated using $M^{(n)}$ and $200$ random masks, but only regarding as positive examples those available at the beginning of iteration $n - 1$, is not statistically significantly higher\footnotemark[\getrefnumber{ftnt_ttest}] than the average MLM loss estimated using $M^{(n - 1)}$ and those same positive examples;
    \item If Step 6 fails, terminate the IPA. Otherwise, pick the top 5 pairs according to $\tilde{C}^{(n)}$, promote them to positive examples in all subsequent iterations, and remove them from the portion of the paired MSA to be optimized.
\end{enumerate}
If several species are present, they are optimized together (see \qnameref{subsec:MLM_loss_meth}) and confidence values from all species are used to select the top 5 pairs.

\subsection*{Pairing based on a single-sequence language model}\label{subsec:methods_esm2}
To assess whether a single-sequence model is able to solve the paralog matching problem, we consider the 650M-parameter version of the model ESM-2~\cite{Lin2022}. We score candidate paired sequences using the MLM loss in \cref{eq:MLM_loss}.
In contrast with MSA Transformer, the input of the model is not paired MSAs but single paired sequences. Therefore, it is sufficient to individually score each possible pair within each species, without needing to consider all permutations. Denoting by $N_k$ the number of sequences from each family in species $k$, the number of possible pairs is $N_k^2$ while the number of permutations is $N_k!$. This complexity reduction allows us to evaluate the scores of all possible pairs. This removes the need of backpropagating the loss on the permutation matrix. Accordingly, this method is much faster, since we only need to use the model in evaluation mode, without gradient backpropagation.

For each candidate paired sequence, we evaluate the average of the MLM losses computed over multiple random masks (with masking probability $p$). Once the average MLM losses are computed for all the $N_k^2$ pairs, we compute the optimal one-to-one matching by using standard algorithms for linear assignment problems \cite{Kuhn55} on the $N_k\times N_k$ matrix containing all the losses.

\subsection*{Assessing the impact of pairing on AFM structure prediction}\label{AFMMeth}

\paragraph{Pairing methods employed in AFM and ColabFold.} When presented with a set of query chains, AFM retrieves homologs of each of the chains by running JackHMMER \cite{Johnson10} on UniProt, and further homology searches on other databases \cite{evans2021protein}. UniProt hits are partitioned into species\footnote{While the official AFM implementation in \url{https://github.com/deepmind/alphafold} uses UniProt ``entry names'' to define species, when possible we instead use NCBI TaxIDs (via UniProt mappings to NCBI tax IDs, retrieved on 17 Dec 2022, corresponding to UniProt Knowledgebase release 2022\_04), which are more accurate.} and ranked within each species by decreasing Hamming distance to the relevant query sequence. A paired MSA is obtained by matching equal-rank hits. Sequences left unpaired are discarded. In addition, AFM produces ``block MSAs'' constructed by ``pairing'' hits from the remaining databases with padding sequences of gaps. The input for AFM comprises the paired MSA and the block MSAs.

While sharing the same architecture and weights as AFM, ColabFold retrieves homologs using MMseqs2 \cite{Steinegger2017} on ColabFoldDB~\cite{mirdita2022colabfold}. In each species, hits are sorted by increasing E-value, and the best hits are paired~\cite{Zeng18,cong2019protein,Green21,Pozzati21,bryant2022improved,mirdita2022colabfold}. Thus, contrary to the default AFM pipeline, the paired MSA in ColabFold contains at most one sequence pair per species for a heterodimer. Because the databases and homology search methods used by ColabFold differ from those used by AFM, a direct comparison does not allow one to isolate the effect of their different pairing schemes. Therefore, we employed the ColabFold pairing method starting from the sequences that are paired in the default AFM pipeline.

\paragraph{Pairing using DiffPALM.} To assess the impact of DiffPALM on complex structure prediction by AFM, we started from the sequences that are paired in the default AFM pipeline. We left out species with large unbalances between the number of sequences in the two families considered. Specifically, if the ratio of the larger to the smaller of these two numbers exceeds an ad-hoc ``maximum size ratio'' MSR (see \cref{tab:dataset_pdb}), if there is only one sequence in both families, or if there are more than $50$ sequences in at least one family, then we do not attempt pairing via DiffPALM, and revert to default AFM pairing. When the full MSA to be paired with DiffPALM is sufficiently deep and/or long, optimizing it as a whole is not possible due to GPU memory limitations. Instead, we partition it into several small enough sub-MSAs, which we optimize independently. We always use the query sequences as positive examples.

\section*{Acknowledgments}
The authors thank Sergey Ovchinnikov for valuable feedback on a preliminary version of this manuscript, and Alex Hawkins-Hooker for interesting conversations.
This project has received funding from the European Research Council (ERC) under the European Union’s Horizon 2020 research and innovation programme (grant agreement No.~851173, to A.-F.~B.).

\section*{Data availability statement}

A Python implementation of DiffPALM is freely available in our GitHub repository: \url{https://github.com/Bitbol-Lab/DiffPALM}.



\newpage
\appendix

\begin{center}
\LARGE{\textbf{Supplementary material}}
\end{center}
\vspace{0.2cm}

\renewcommand{\thesection}{S\arabic{section}}
\renewcommand{\thefigure}{S\arabic{figure}}
\setcounter{figure}{0}
\renewcommand{\thetable}{S\arabic{table}}
\setcounter{table}{0}
\renewcommand{\theequation}{S\arabic{equation}}
\setcounter{equation}{0}

\section{Supplementary methods}

\subsection{MSA Transformer and masked language modeling for MSAs}
\label{supp_meth:MSA-Tr_MLM}

We use the MSA Transformer model \cite{rao2021msa}, which takes MSAs as inputs and was trained with a variant of the masked language modeling (MLM) objective \cite{devlin2019bert} on a training set of 26 million MSAs constructed from UniRef50 clusters.
The model's training objective was to correctly predict the identity of randomly masked residue positions in the MSAs in its training set.
Specifically, it was trained to minimize an MLM loss, which reads, for an MSA $\mathcal{M}$, and its masked version $\widetilde{\mathcal{M}}$:
\begin{equation}
\label{eq:MLM_loss}
    \mathcal{L}_{\text{\tiny{MLM}}}(\mathcal{M}, \widetilde{\mathcal{M}}; \theta) = - \sum_{\mathclap{(m,i)\, \in\, \text{mask}}} \log p(x_{m,i} \, | \, \widetilde{\mathcal{M}};\theta) \, .
\end{equation}
Here, $x_{m,i}$ denotes the amino acid at the $i$-th residue position (column) in the $m$-th sequence (row) of $\mathcal{M}$, while $\theta$ stands for all the model parameters.
At each residue position in the input MSA, MSA Transformer outputs a probability for each of the 21 possible amino-acid and gap symbols, and $p(x_{m,i}\, | \,\widetilde{\mathcal{M}}; \theta)$ in \cref{eq:MLM_loss} is the probability associated with the correct residue $x_{m, i}$ at MSA position $(m, i)$.
MSA Transformer's architecture interleaves multi-headed (tied) row attention blocks and (untied) column attention blocks, over several layers. Therefore, the accessible context for a masked residue consists not only of amino acids at different positions along the same sequence, but also of amino acids from other sequences \cite{rao2021msa,lupo2022protein}. This allows the model to capture coevolution information. After pre-training, each term in the right-hand side of \cref{eq:MLM_loss} can be interpreted as the model's estimate of the (negative) log-likelihood of the amino acid $x_{m, i}$ at a masked position $(m, i)$~\cite{Wang19,goyal2021exposing,rao2021transformer}.

\subsection{A differentiable formulation of paralog matching}
\label{supp_meth:diff_matching}

We formulate a differentiable optimization problem that can be more efficiently solved than the brute-force search, using gradient methods. The goal is to obtain sets of within-species matchings (and thus permutations) that minimize our MLM loss.

The set $\mathcal{P}_N$ of permutation matrices of $N$ objects can be parameterized exactly by square matrices $X$ via the \emph{matching operator}
\begin{equation}
\label{eq:matching_op}
  M(X) = \argmax_{P \in \mathcal{P}_N} [\mathrm{trace}(P^{\mathrm{T}} X)],
\end{equation}
which can be computed using standard non-differentiable algorithms for linear assignment problems \cite{Kuhn55}.\footnote{More precisely, the right-hand side of \cref{eq:matching_op} has a unique solution for almost all $X$ \cite{Mena2018}.}

We exploit the fact, shown in \cite{Mena2018}, that permutation matrices can be approximated arbitrarily well by using the \emph{Sinkhorn operator} $S$, which is defined on square matrices $X$ as follows:
\begin{equation}
\label{eq:sinkhorn_op}
    S(X) = \lim_{l \to \infty} S^{l}(X), \quad \text{where} \quad S^{l}(X) = (\mathcal{T}_\mathrm{c} \circ \mathcal{T}_\mathrm{r})^l(\exp(X)),
\end{equation}
$\mathcal{T}_\mathrm{c}$ and $\mathcal{T}_\mathrm{r}$ are the row- and column-wise normalization operators, and $\exp$ denotes the component-wise matrix exponential.\footnote{That is, $S^{l}$ consists of applying $\exp$ and then iteratively normalizing rows and columns $l$ times.}
More precisely, $M(X) = \lim_{\tau \to 0^+} S(X / \tau)$ for almost all $X$ \cite[Theorem 1]{Mena2018}.
Hence, by choosing a suitably small value of $\tau$, and using $S^l$ [\cref{eq:sinkhorn_op}] instead of $S$ for a suitably large $l$, we can define a smooth mapping $\hat{S}(X) = S^l(X/\tau)$ which sends arbitrary square matrices to ``soft permutations'' approximating \emph{bona fide} (``hard'') permutations.
In practice, we use $\tau = 1$ and $l = 10$.

Applying general soft permutations directly on an MSA (after one-hot encoding its residues) yields a dataset consisting of ``amino acid mixtures'' at each MSA position.
Such datasets are out of distribution relative to MSA Transformer's pre-training since it was trained on single amino acid embeddings, not mixtures of them.
Besides, we wish to optimize for an MLM loss defined on realistic MSAs.
Therefore, in order to be able to backpropagate through $\hat{S}$, while also evaluating MLM losses only on MSAs shuffled by hard permutations, we compute the full matching operator $M$ [\cref{eq:matching_op}] in the forward pass, but propagate gradients backwards through $\hat{S}$ alone.\footnote{See \cite{norn2021protein} for a similar use of ``gradient bypassing'' in the context of protein design.
We write the hard permutation as $[M(X) - \hat{S}(X)] + \hat{S}(X)$, and halt gradient backpropagation through the term in square brackets.}

\subsection{Datasets}
\label{subsec:Datasets}

\paragraph*{Benchmark prokaryotic datasets.}
We developed and tested DiffPALM using joint MSAs extracted from two datasets.
The first dataset is composed of 23,632 cognate pairs of histidine kinases (HK) and response regulators (RR) from the P2CS database~\cite{Barakat09,Barakat11}, paired using genome proximity, and previously described in~\cite{Bitbol16,Bitbol18}.
The average size of the species in this dataset is $10.23$ (standard deviation: $7.85$).

The second dataset consists of 17,950 ABC transporter protein pairs, homologous to the \textit{Escherichia coli} MALG-MALK pair of maltose and maltodextrin transporters, also paired using genome proximity~\cite{Ovchinnikov14,Bitbol16}.
The average size of the species in this dataset is $5.68$ (standard deviation: $5.60$).

Out of each of these two benchmark datasets of known interacting pairs, we consider paired MSAs of depth $\sim 50$, constructed by selecting random species from the full dataset. Specifically, species are added one by one until an MSA depth between $45$ and $55$ is reached. For such shallow MSAs, existing coevolution-based methods do not perform well. Note also that MSA Transformer's large memory footprint constrains the depth and length of input MSAs.

\paragraph*{Eukaryotic complexes.} \label{supp_datasets_eukaryotic}

We considered targets whose structure are not in the training set of AFM with v2 weights, and where default AFM predictions are poor.
Specifically, we started from those eukaryotic targets from Table A1 of \cite{Chen23} and from the ``Benchmark 2'' dataset in \cite{evans2021protein} whose PDB structures were released after the training cutoff for the AFM v2 weights (April 30, 2018).
Among those, we focused on multimers with no more than 2 different types of monomers, where both monomers come from the same species, and with paired sequences not longer than $500$ amino acids, due to GPU memory constraints.
Finally, we further restricted to the 15 targets with default AFM predictions yielding the poorest reported DockQ score. They are listed in \cref{tab:dataset_pdb}. All of them are heterodimers, except 6ABO which is a heterotetramer complex made of two IFFO1 and two XRCC4 molecules.

\subsection{General points on AFM} \label{subsec:generalities_AFM}

For all structure prediction tasks, we use the five pre-trained AFM models with v2 weights \cite{evans2021protein}. We use full genomic databases and code from release v2.3.1 of the official implementation in \url{https://github.com/deepmind/alphafold}. We use no structural templates, and perform 3 recycles for each structure, without early stopping. We relax all models using AMBER.

When using all pairing methods (default AFM, DiffPALM, orthology-based), we also retained the block MSAs retrieved by the default AFM pipeline~\cite{bryant2022improved}.

The AFM confidence score is defined as $0.8 \cdot \mathrm{iptm} + 0.2 \cdot \mathrm{ptm}$, where iptm is the predicted TM-score in the interface, and ptm the predicted TM-score of the entire complex~\cite{evans2021protein}.


\section{Supplementary tables}

\begin{table}[htbp]
    \centering
    \begin{tabular}{cccccccccccc}
        \toprule
PDB ID	&	$L_A$	&	$L_B$	& $D$ &	$F_\mathrm{paired}$	&	$\langle d_p\rangle$ & MSR	&	$F_\mathrm{same}$	&	$F_\mathrm{pred}$ & $D_\mathrm{DiffPALM}$ & $D_\mathrm{eff}^A$ & $D_\mathrm{eff}^B$ \\
\midrule
6QU1	&	322	&	48	&	9267	&	0.02	&	3.4	&	5	&	0.37	&	0.45	& 76 & 20 & 11 \\
6POG	&	114	&	249	&	18390	&	0.20	&	30.1	&	3	&	0.03	&	0.22	& 821  & 114 & 93 \\
6THL	&	240	&	185	&	3515	&	0.03	&	2.2	&	5	&	0.40	&	0.46	& 53 & 28 & 26 \\
6L5K	&	98	&	113	&	15857	&	0.29	&	22.9	&	3	&	0.05	&	0.75	& 3478 & 402 & 644 \\
6A6I	&	98	&	76	&	4793	&	0.11	&	3.1	&	5	&	0.30	&	0.48	& 259 & 57 & 87 \\
5Z5K	&	380	&	66	&	6349	&	0.03	&	12	&	10	&	0	&	0.02	& 3 & 3 & 3 \\
6FYH	&	124	&	76	&	15285	&	0.51	&	7.1	&	3	&	0.14	&	0.72	& 5550 & 1703 & 1640 \\
6WCW	&	184	&	254	&	20435	&	0.21	&	6.7	&	3	&	0.18	&	0.57	& 2509 & 263 & 402 \\
5XLN	&	190	&	45	&	9434	&	0.04	&	3.4	&	5	&	0.30	&	0.65	& 228 & 30 & 31 \\
7BQU	&	114	&	28	&	5915	&	0.04	&	2.9	&	5	&	0.44	&	0.65	& 152 & 67 & 61 \\
6ABO	&	227	&	82	&	5058	&	0.10	&	4.5	&	3	&	0.32	&	0.56	& 310 & 65 & 25 \\
6INE	&	267	&	177	&	19227	&	0.18	&	4	&	3	&	0.26	&	0.39	& 1334 & 402 & 480 \\
6IRE	&	234	&	194	&	7599	&	0.12	&	4.2	&	3	&	0.21	&	0.63	& 591 & 120 & 210 \\
6PNQ	&	202	&	168	&	5694	&	0.20	&	8.7	&	5	&	0.13	&	0.25	& 282 & 117 & 23 \\
6GK2	&	106	&	92	&	3062	&	0.08	&	2.4	&	5	&	0.35	&	0.59	& 145 & 20 & 40 \\
        \bottomrule
    \end{tabular}
\caption{\textbf{Dataset of eukaryotic complexes.} All 15 eukaryotic protein complexes considered here are listed by their PDB ID, and various MSA properties are given. $L_A$ and $L_B$ are the lengths of the aligned amino acid sequences of the two chains A and B considered. $D$ denotes the depth of the full MSA built by AFM, consisting of the paired MSA and the block MSAs (see \qnameref{AFMMeth}). $F_\mathrm{paired}$ is the fraction of sequences that are paired by AFM, i.e.\ the depth of the paired MSA divided by $D$. $\langle d_p\rangle$ is the average depth of the MSAs of the single species in the AFM-paired MSA. Thus it is the average of the largest depth among the two chains, since the other one is completed by padding sequences of gaps.  MSR stands for ``maximum size ratio'': if the ratio of the larger to the smaller of the depths of MSAs A and B is larger than MSR, species are not paired by DiffPALM. $F_\mathrm{same}$ is the fraction of pairs predicted by DiffPALM that is identical to the pairs predicted by the default AFM pairing method. $F_\mathrm{pred}$ is the ratio of the number of pairs predicted with DiffPALM to the number predicted using the default AFM pairing method. $D_\mathrm{DiffPALM}=D\times F_\mathrm{paired}\times F_\mathrm{pred}$ denotes the depth of the MSA which is paired by DiffPALM. $D_\mathrm{eff}^A$ (resp.\ $D_\mathrm{eff}^B$) is the effective depth corrected with phylogenetic weights (with Hamming distance threshold $0.2$)~\cite{Weigt09} of the MSA associated to chain A (resp.\ chain B) in the MSAs which are paired by DiffPALM. These effective depths quantify MSA diversity. Rows are ordered by increasing mean DockQ score for the default AFM pairing method (cf.\ \cref{fig-sup:results_trajectories}).}
\label{tab:dataset_pdb}
\end{table}

\begin{table}
\centering
\begin{tabular}{cccccc}
        \toprule
        MSAs & Pairing method & Pos.\ Ex. & $N_\mathrm{runs}$ & Precision-100 & Precision-10 \\
        \midrule
        HK-RR & Chance & - & - & 0.09 & - \\
        HK-RR & DCA-IPA \cite{Bitbol16} & 0 & - & 0.16 & - \\
        HK-RR & MI-IPA \cite{Bitbol18} & 0 & - & 0.15 & - \\
        HK-RR & GA-IPA \cite{Gandarilla23} & 0 & - & 0.18 & - \\
        HK-RR & ESM-2 (650M) & 0 & - & 0.11 -- 0.16 & - \\
        HK-RR & DiffPALM-MRA & 0 & 5 & 0.37 & 0.67 \\
        HK-RR & DiffPALM-MRA & 0 & 20 & 0.39 & 0.71 \\
        HK-RR & DiffPALM-IPA & 0 & $20 + 10$ & 0.44 & - \\
        HK-RR & DiffPALM-MRA & 11 & 5 & 0.51 & 0.87 \\
        HK-RR & DiffPALM-MRA & 19 & 5 & 0.61 & 0.95 \\
        HK-RR & DiffPALM-MRA & 45 & 5 & 0.74 & 0.99 \\
        \midrule
        MALG-MALK & Chance & - & - & 0.20 & - \\
        MALG-MALK & DCA-IPA \cite{Bitbol16} & 0 & - & 0.31 & - \\
        MALG-MALK & MI-IPA \cite{Bitbol18} & 0 & - & 0.32 & - \\
        MALG-MALK & GA-IPA \cite{Gandarilla23} & 0 & - & 0.42 & - \\
        MALG-MALK & ESM-2 (650M) & 0 & - & 0.19 -- 0.31 & - \\
        MALG-MALK & DiffPALM-MRA & 0 & 5 & 0.55 & 0.84 \\
        \bottomrule
    \end{tabular}
    \caption{\textbf{Performance of pairing by DiffPALM vs.\ baselines.} We report the pairing precision for variants of DiffPALM, namely MRA \& IPA with various numbers of positive examples (Pos.\ Ex.) and runs ($N_\mathrm{runs}$), as well as for different baseline methods, on $40$ MSAs comprising about 50 HK-RR or MALG-MALK pairs. With all methods, a full one-to-one within-species pairing is produced for each MSA, and performance is measured by precision, namely, the fraction of correct pairs among predicted pairs, averaged over the 40 MSAs considered. In ``precision-100'', this fraction is computed over all predicted pairs (100\% of them). In ``precision-10'', it is calculated over the top $10 \%$ predicted pairs, when ranked by predicted confidence.
    For the IPA method, we use 20 runs of MRA as starting point, and we add 5 fixed pairs at each new run, see \qnameref{subsec:algo}. The chance expectation, and the performance of DCA-IPA \cite{Bitbol16}, MI-IPA \cite{Bitbol18}, and GA-IPA \cite{Gandarilla23} are reported as baselines. We also consider a pairing method based on the scores given by the ESM-2 (650M) protein language model \cite{Lin2022}, see \qnameref{subsec:methods_esm2}. For this method, we consider 10 different values of masking probability from $0.1$ to $1.0$, and we report the range of precisions obtained.}
    \label{tab:all-data}
\end{table}

\newpage

\section{Supplementary figures}

\begin{figure}[!h]
    \centering
    \includegraphics[width=0.9\textwidth]{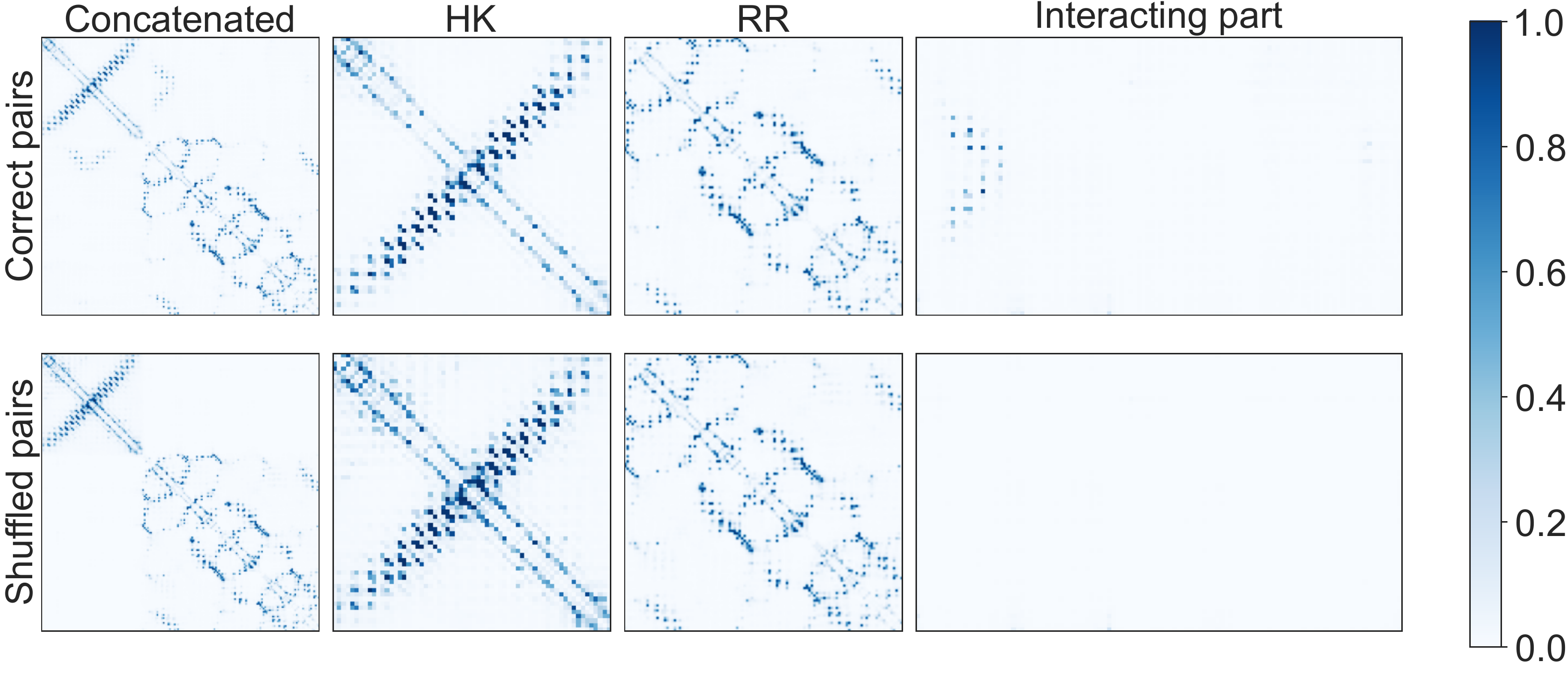}
    \caption{Comparison of contact maps predicted by MSA Transformer for the correct pairing of an HK MSA and an RR MSA (``Correct pairs''), and for an incorrect pairing (``Shuffled pairs''). We observe that MSA Transformer is able to correctly predict the inter-protein contacts when given as input a paired MSA made of correctly matched sequences. Conversely, if the model is given as input a paired MSAs where rows have been shuffled before pairing, it is not able to recover the inter-protein contact map (even though it correctly recovers correctly the intra-protein contact maps). These results suggests that MSA Transformer can distinguish between interacting and non-interacting pairs of protein sequences, despite the fact that dimers or paired MSAs were not in the training set used for its MLM pre-training \cite{rao2021msa}.}
    \label{fig-sup:concat_contacts}
\end{figure}

\begin{figure}[t!]
    \centering
    \includegraphics[width=\textwidth]{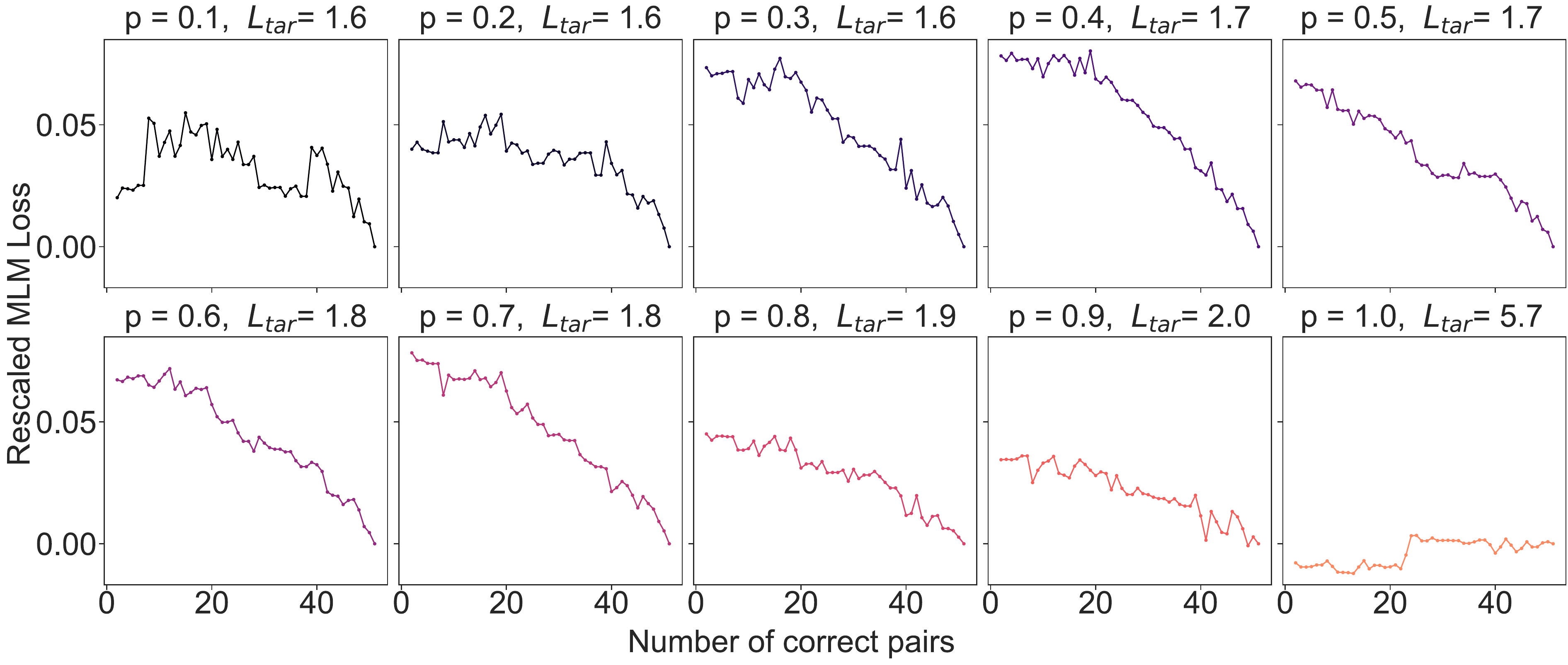}
    \caption{\textbf{MLM loss vs.\ number of correct pairs for different masking probabilities.}
    We use an MSA of $M=50$ sequences and 5 different species.
    To estimate the expected loss accurately, we used 20 different masks at each step.
    $L_{tar}$ denotes the expected loss when all pairs are correctly matched.
    For visualization purposes, in every plot we rescale the loss by shifting it by $L_{tar}$. We find that our MLM loss in \cref{eq:MLM_loss} decreases for increasing numbers of correctly matched sequences in the MSA. We see that the sweet spot of the masking probability $p$ (i.e.\ the value that gives steeper and smoother loss curves) is at moderately high values ($0.4 \leq p \leq 0.7$). As explained in \qnameref{sec:methods}, high masking probabilities make it more challenging for the model to predict the masked amino acids using only information coming from the masked MSA, thus encouraging it to use, instead, information coming from the matched MSA. This motivates our choice of a masking probability of $p = 0.7$.}
    \label{fig-sup:loss_vs_pairs}
\end{figure}

\begin{figure}[h!]
\centering
\includegraphics[width=0.8\textwidth]{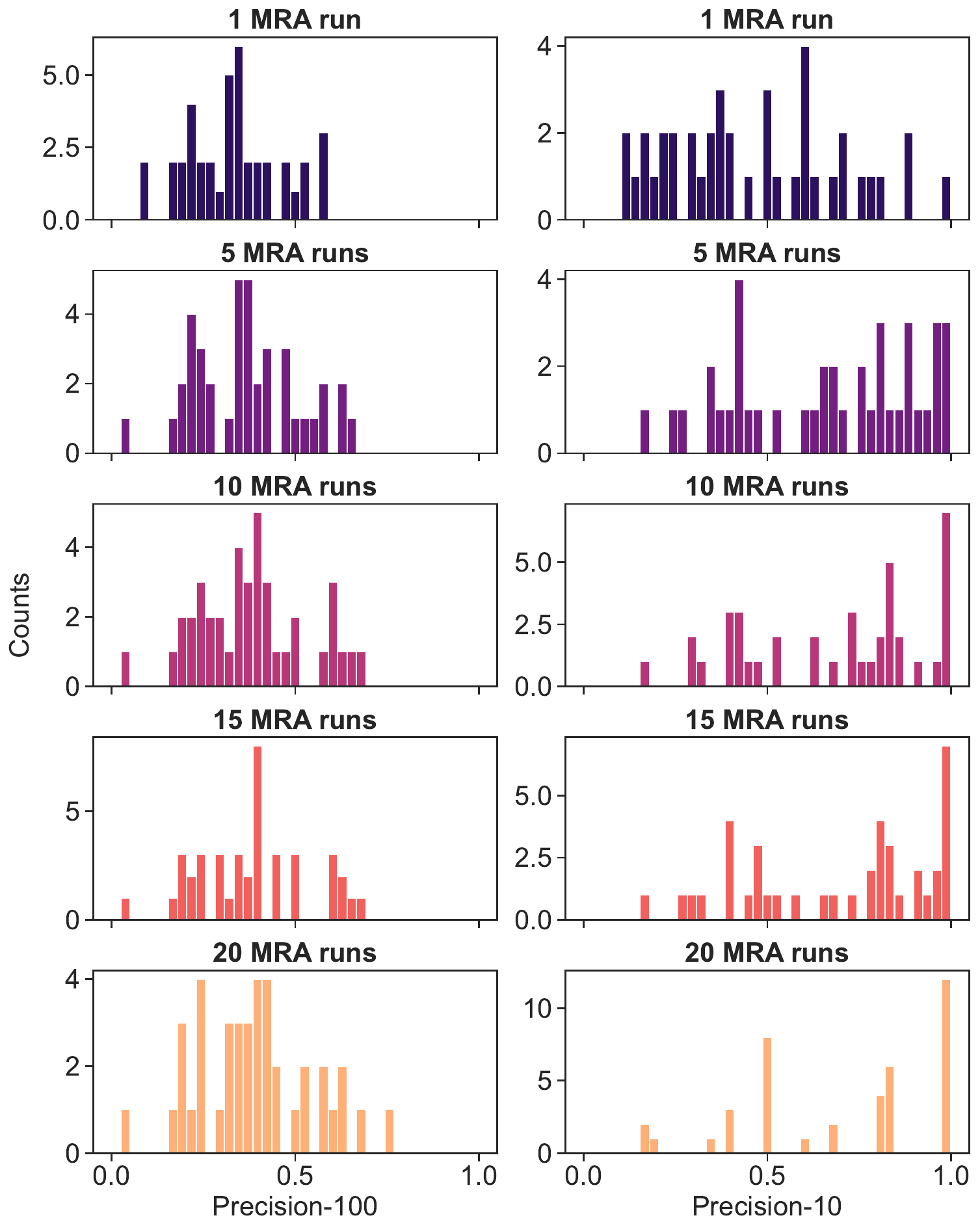}
\caption{\textbf{Distribution of precision scores for different number of MRA runs.} We report the histograms of precision scores obtained by the MRA variant of DiffPALM on each of the 40 MSAs comprising about 50 HK-RR pairs (used in \qnameref{subsec:results_prokaryotic}), for different number of MRA runs. Precision-100 and precision-10 are defined in \cref{fig:results_mra} and \cref{tab:all-data}. We observe a skewed distribution for precision-10 scores, especially after many MRA runs: a very high precision is reached for many MSAs, but low precisions are obtained for some. \cref{fig:results_mra} displays the average and the standard error of each of the distributions shown here.} 
\label{fig-sup:results_distributions}
\end{figure}

\begin{figure}[htbp]
\centering
\includegraphics[width=\textwidth]{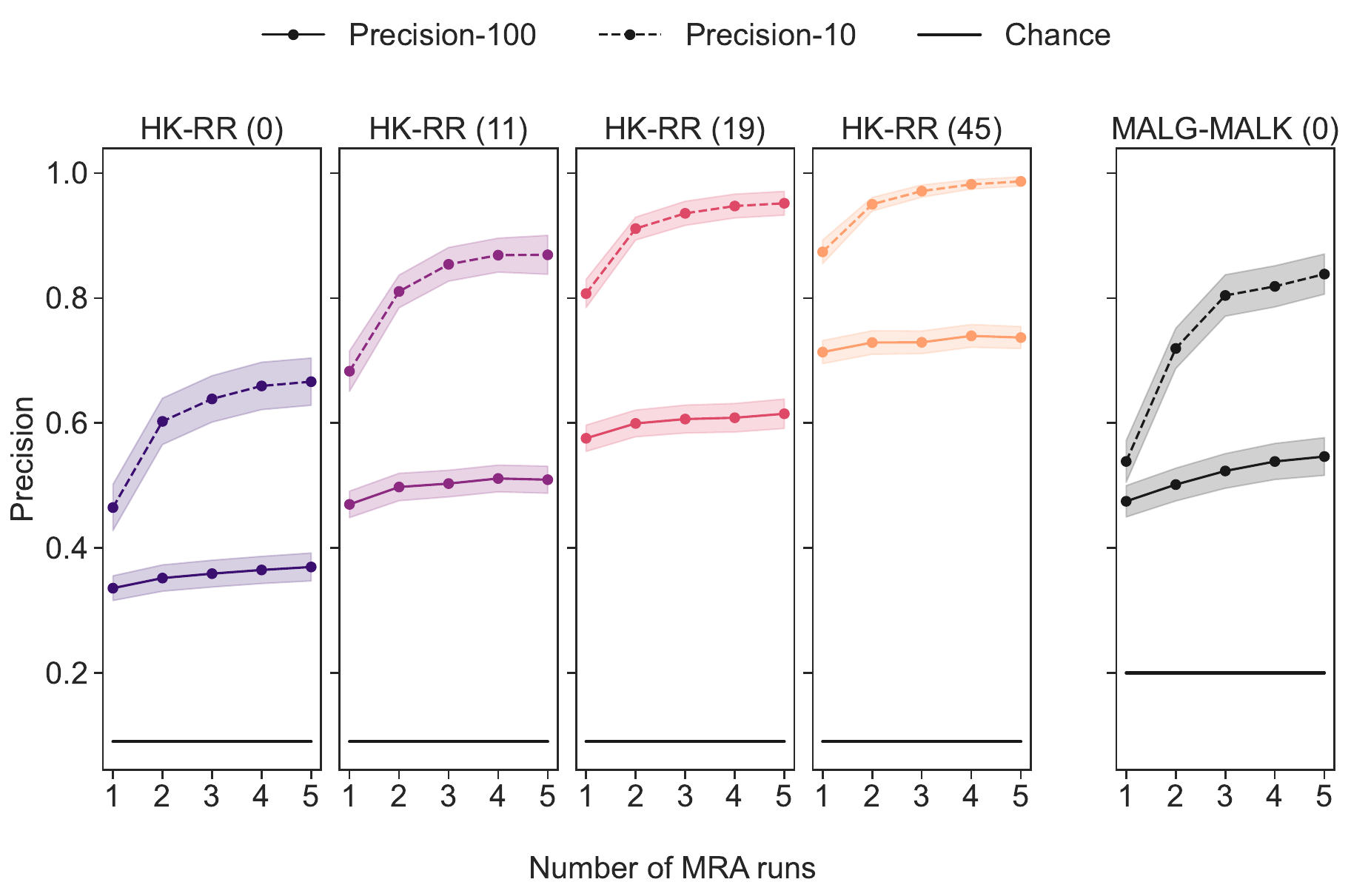}
\caption{\textbf{Performance of DiffPALM  for different numbers of positive examples and for two pairs of protein families.} The performance of DiffPALM is plotted versus the number of MRA runs. As in \cref{fig:results_bars}, we compare runs with various numbers of positive examples (left panels) on the 40 MSAs comprising about 50 HK-RR pairs (used in \qnameref{subsec:results_prokaryotic}), and runs on 40 MSAs comprising about 50 MALG-MALK pairs with no positive example (right panel). Precision-100 and precision-10 are defined in \cref{fig:results_mra} and \cref{tab:all-data}. The protein families considered and the number of positive examples are indicated in the title of each panel (the latter between brackets). In each case, we plot the mean value over the 40 different
MSAs considered and the standard error interval. The chance expectation is shown for reference.}
\label{fig-sup:results_iters}
\end{figure}

\begin{figure}[htbp]
\centering
\includegraphics[width=\textwidth]{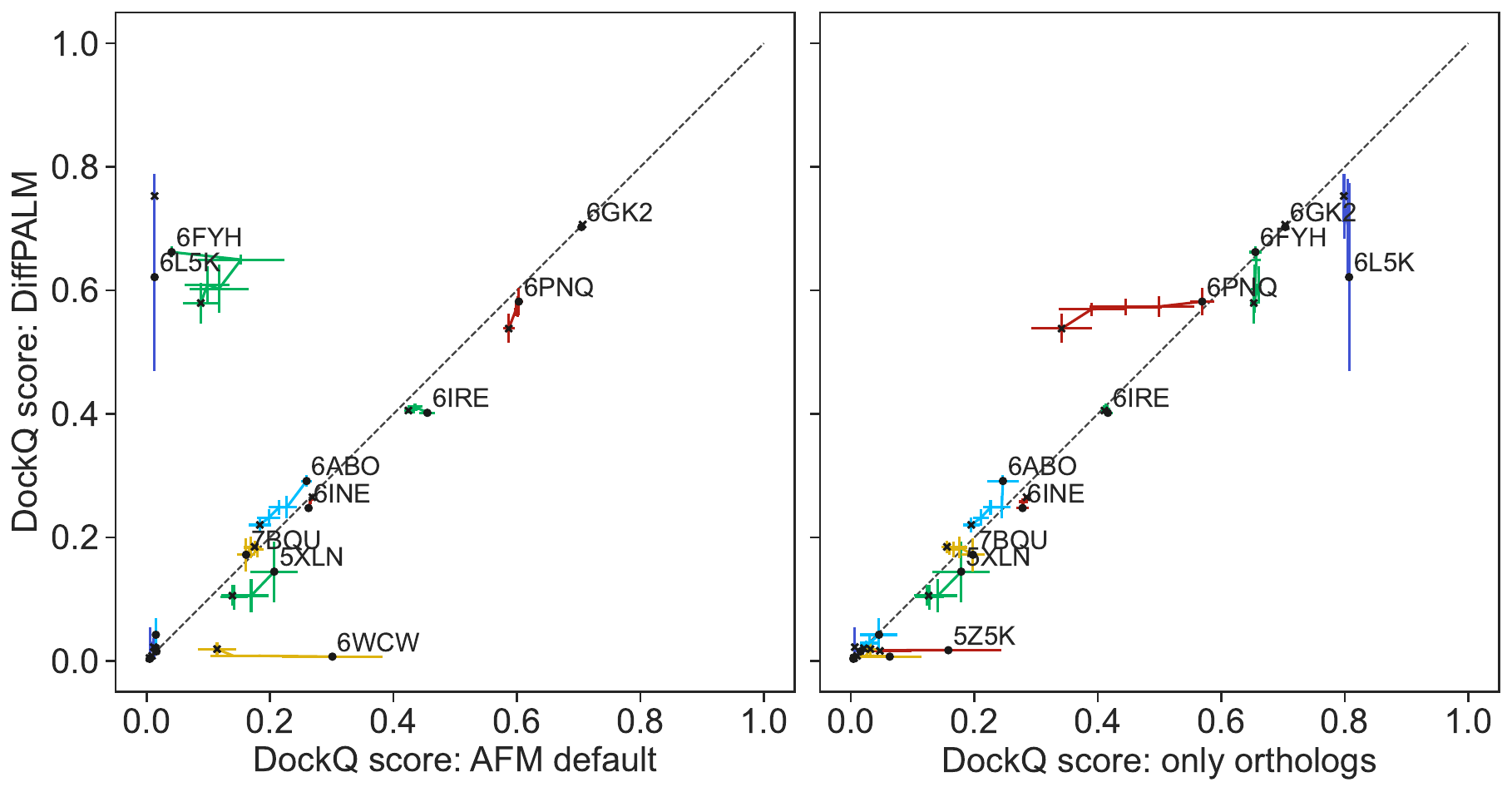}
\caption{\textbf{Performance of structure prediction by AFM using different MSA pairing methods.} We report the performance of AFM, in terms of DockQ scores, for the 15 complexes listed in \cref{tab:dataset_pdb}, using three different pairing methods on the same initial unpaired MSAs. Left panel: DiffPALM versus default AFM pairing method. Right panel: DiffPALM versus only pairing orthologs to the two query sequences. As in \cref{fig:results_afm_swarm}, for each complex, AFM is run five times, and the five top predicted structures by AFM confidence are considered each time, yielding 25 predicted structures total. For each complex, we show ``trajectories" of performance starting from the top-confidence predicted structure (black circular marker) and ending with all predicted structures up to and including the fifth one (black cross marker). Results are averaged over the 5 runs and standard errors are shown as error bars. Points with DockQ below $0.1$ are not labelled with their PDB ID for graphical reasons.
Note that \cref{fig:results_afm_swarm} restricts to those complexes where any two of these three pairing methods yield a significant difference ($>10\%$) in average DockQ scores, among those shown here and listed in \cref{tab:dataset_pdb}. 
} 
\label{fig-sup:results_trajectories}
\end{figure}

\clearpage

\begin{figure}[htbp]
    \centering
    \begin{subfigure}[t]{0.49\textwidth}
        \centering
        \includegraphics[width=\linewidth]{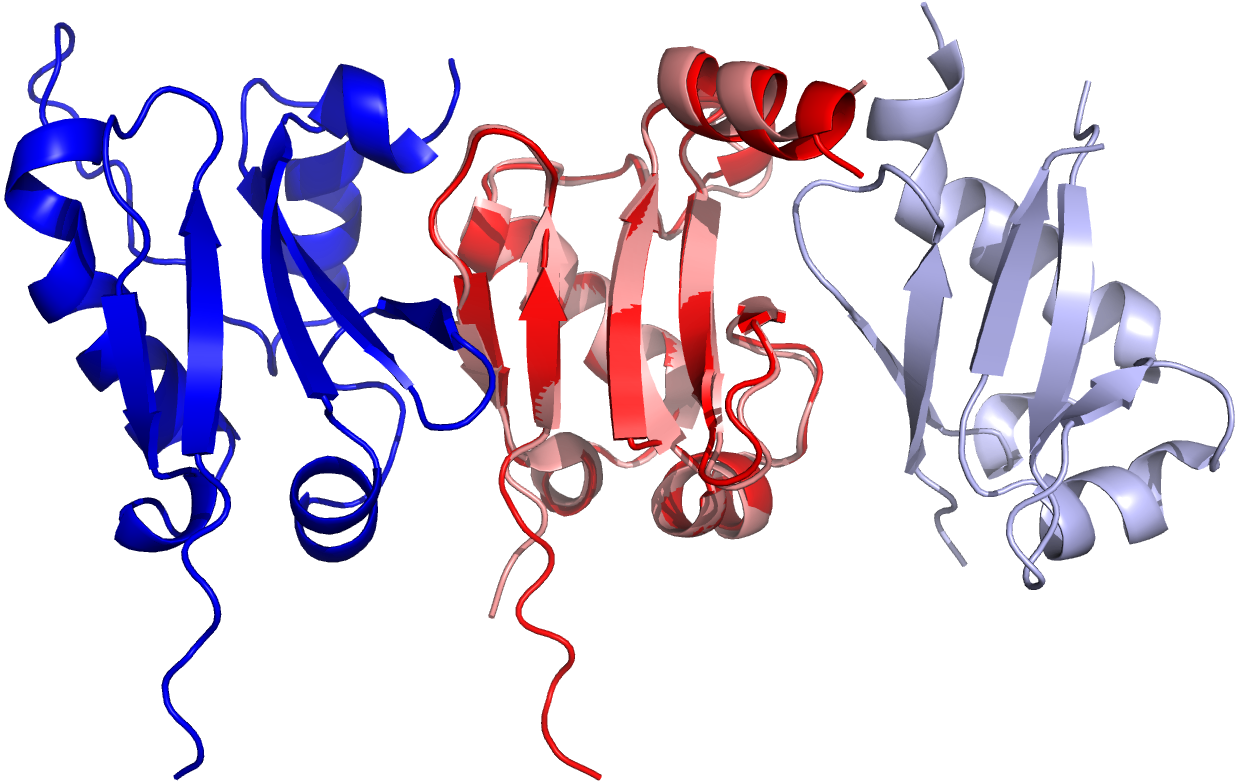} 
        \caption{AFM default}
    \end{subfigure}
    \hfill
    \begin{subfigure}[t]{0.35\textwidth}
        \centering
        \includegraphics[width=\linewidth]{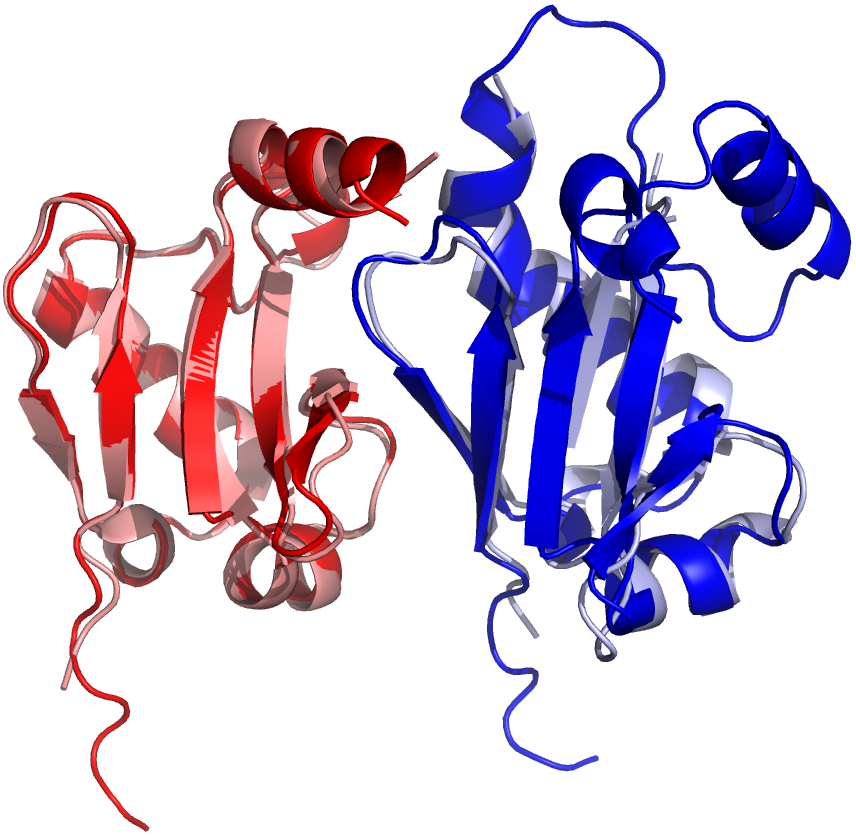} 
        \caption{DiffPALM}
    \end{subfigure}
    \caption{\textbf{Comparing the AFM default MSA pairing strategy with DiffPALM, for structure 6L5K.} In both panels, we superimpose the experimental structure of 6L5K with a structure predicted using AFM. Chains A and B of the PDB structure are colored in salmon and light blue respectively, while chains A and B of both predicted structures are colored in bright red and bright blue respectively. (a) Comparing the experimental structure with a typical high-confidence prediction generated with the default MSA pairing pipeline. (b) Comparing the experimental structure with a typical high-confidence prediction generated with our MSA pairing pipeline based on DiffPALM.}
    \label{fig-sup:6l5k_pymol}
\end{figure}

\vspace{1cm}

\begin{figure}[htbp]
    \centering
    \begin{subfigure}[t]{0.45\textwidth}
        \centering
        \includegraphics[width=\linewidth]{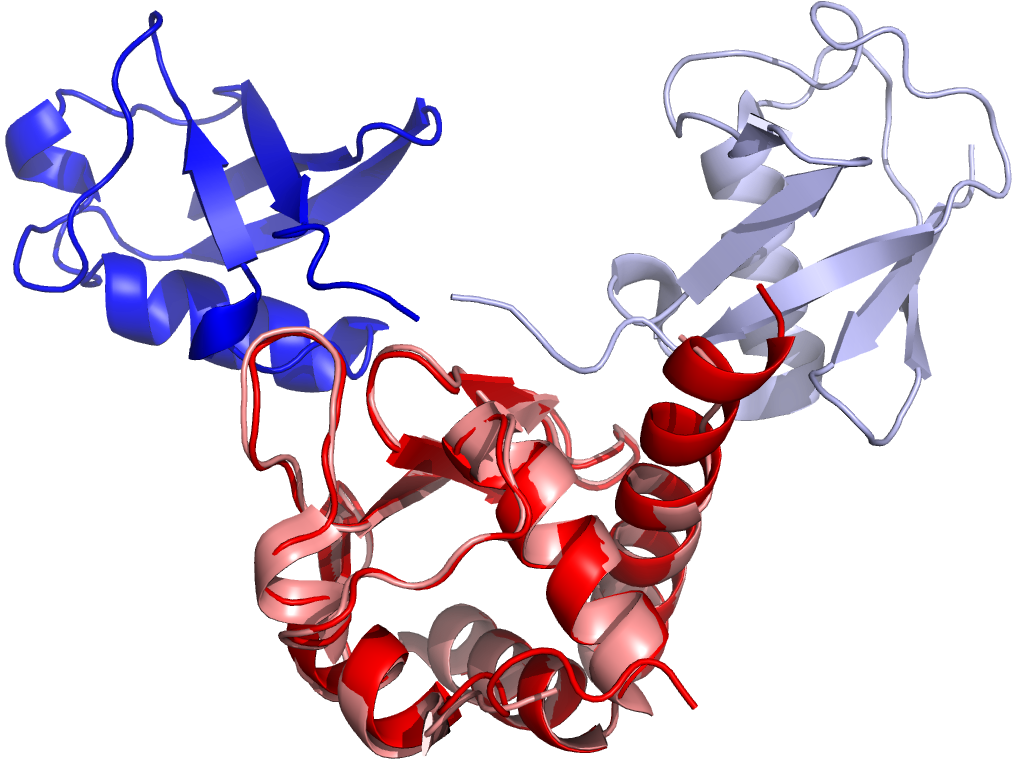} 
        \caption{AFM default}
    \end{subfigure}
    \hfill
    \begin{subfigure}[t]{0.365\textwidth}
        \centering
        \includegraphics[width=\linewidth]{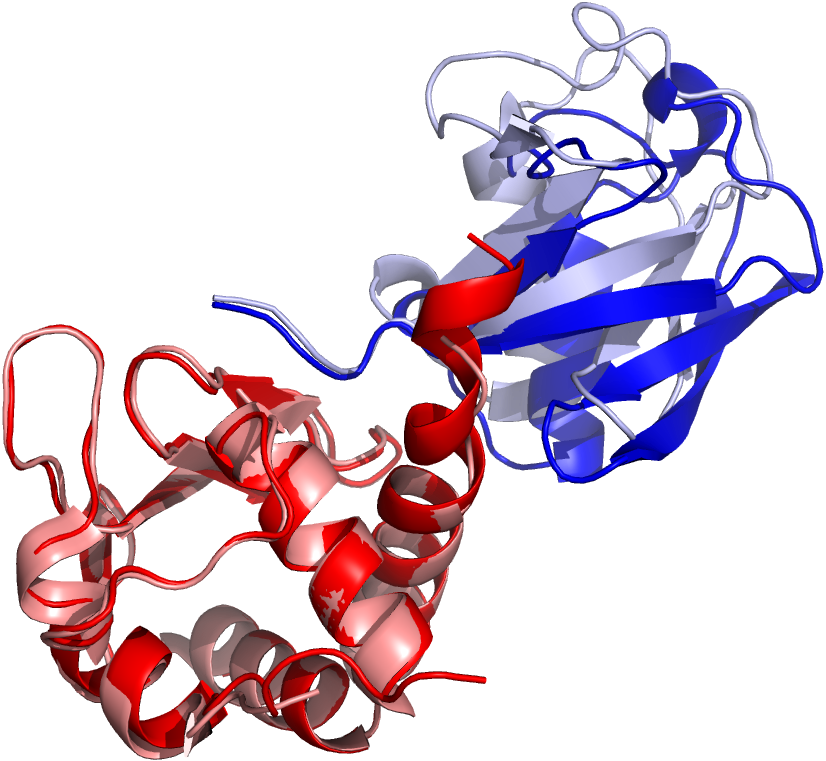} 
        \caption{DiffPALM}
    \end{subfigure}
    \caption{\textbf{Comparing the AFM default MSA pairing strategy with DiffPALM, for structure 6FYH.} Same as \cref{fig-sup:6l5k_pymol}, but for 6FYH.}
    \label{fig-sup:6fyh_pymol}
\end{figure}

\clearpage

\begin{figure}[htbp]
\centering
\includegraphics[width=\textwidth]{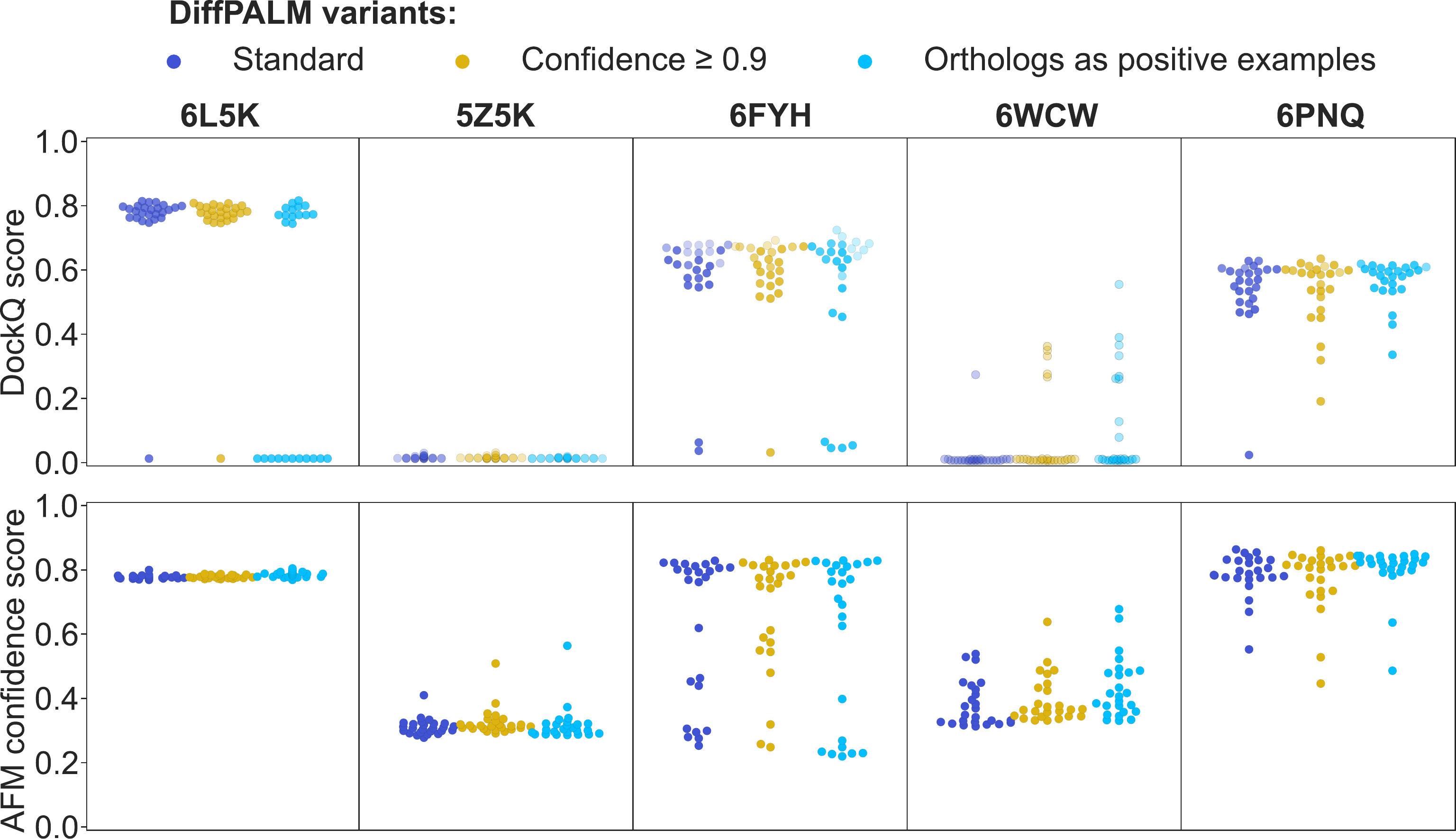}
\caption{\textbf{Performance of AFM using different variants of DiffPALM.} Same as \cref{fig:results_afm_swarm}, but here we compare the standard DiffPALM method with two of its variants: one where we only use pairs with high predicted confidence ($\geq 0.9$) as input to the AFM pipeline, and one where we use orthology-based pairs (i.e.\ those employed in the ``Only orthologs" case shown in \cref{fig:results_afm_swarm}) as positive examples for DiffPALM, and use the pairs predicted by DiffPALM, as well as the positive examples, as input of AFM.
}
\label{fig-sup:results_afm_swarm}
\end{figure}

\begin{figure}[htbp]
\centering
\includegraphics[width=0.95\textwidth]{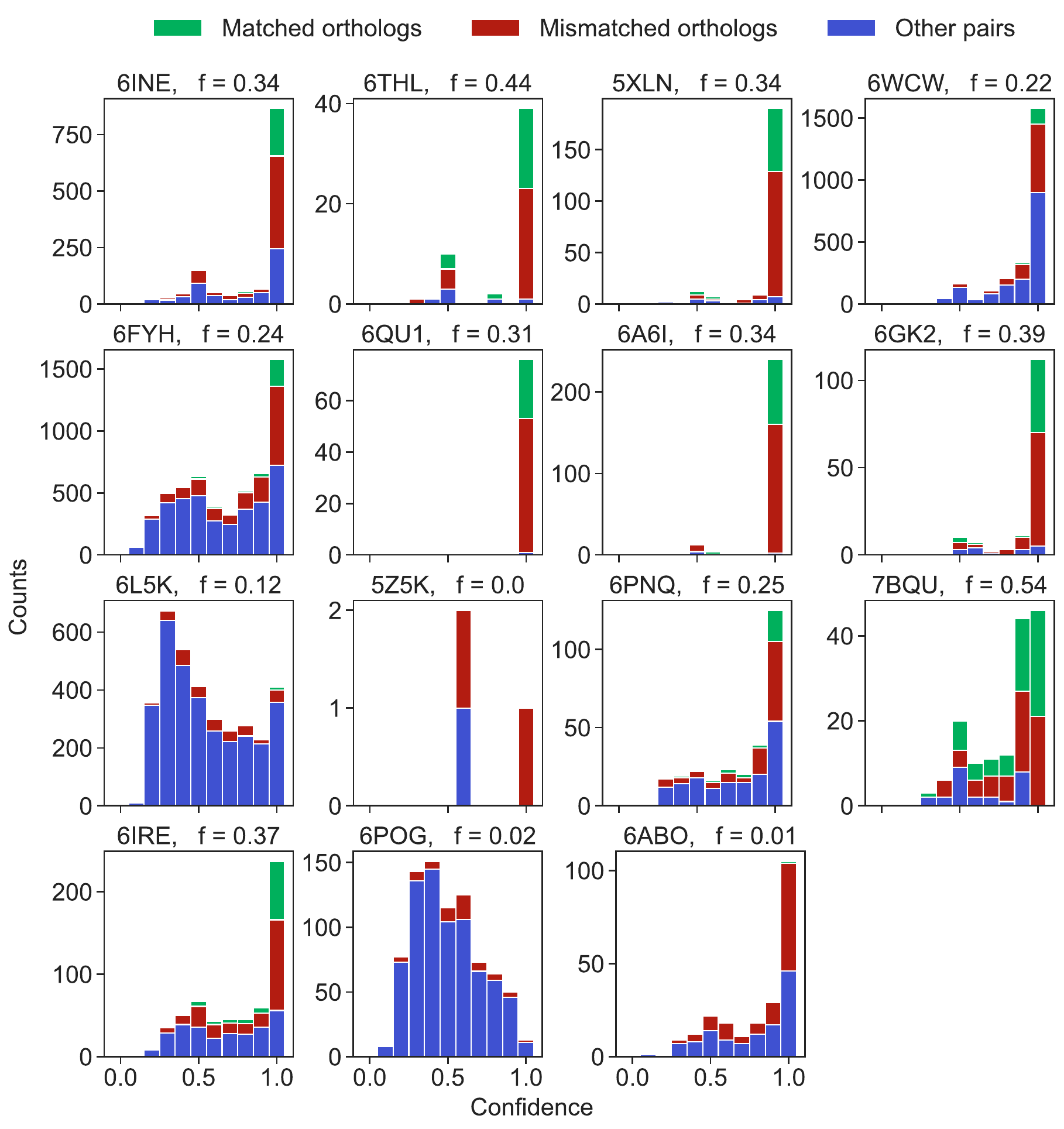}
\caption{\textbf{Confidence of DiffPALM predictions.} We show, for the 15 complexes listed in \cref{tab:dataset_pdb}, histograms of the DiffPALM confidence values (see \qnameref{subsec:confidence}). We distinguish the orthology-based pairs that are recovered by DiffPALM, the otherwise paired orthologs, and all the other paired sequences. We indicate in panel titles the value of the fraction $f$ of orthology-based pairs that are recovered by DiffPALM.}
\label{fig-sup:results_conf}
\end{figure}

\begin{figure}[htbp]
\centering
\includegraphics[width=0.95\textwidth]{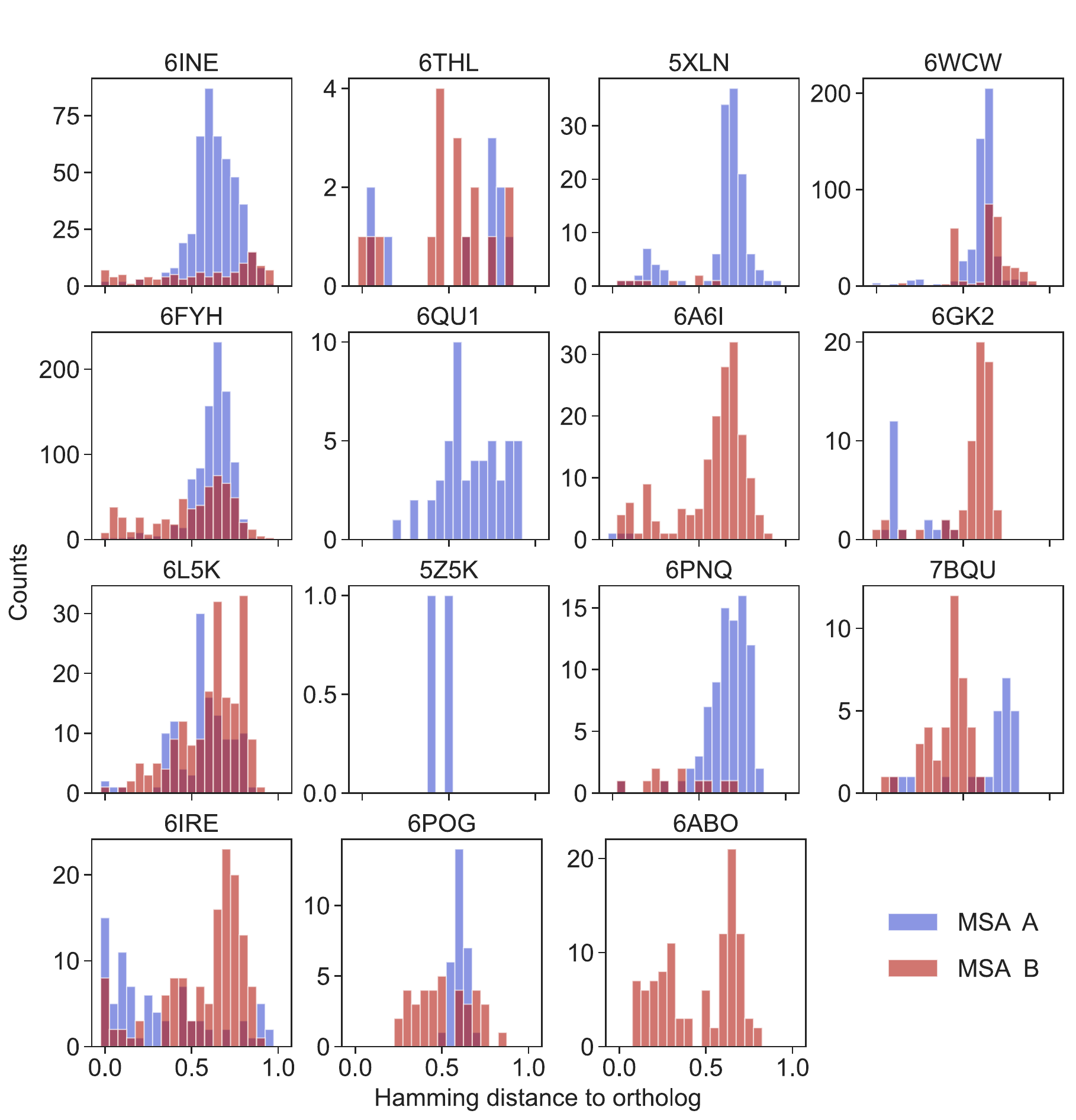}
\caption{\textbf{Hamming distance to the orthologs of all the mismatched pairs predicted by DiffPALM.} We show, for the 15 complexes listed in \cref{tab:dataset_pdb}, histograms of the Hamming distance between the partner predicted by DiffPALM and the one predicted by matching orthologs to the query sequences, whenever they differ. In practice, for each sequence A in family A which is paired with a partner B using orthology, but with a different partner b using DiffPALM, we measure the Hamming distance between B and b. A similar protocol is conducted for each sequence B in family B. These distances allow us to compare the pairs predicted by DiffPALM to the orthology-based pairs. Note that the total counts of the distributions regarding MSA A and MSA B generally differ. This happens because DiffPALM might pair orthologs to padding sequences of gaps: in this case, we do not report Hamming distances.}
\label{fig-sup:results_ham}
\end{figure}

\end{document}